\documentclass[fleqn,usenatbib]{mnras}

\usepackage{newtxtext,newtxmath}

\usepackage[T1]{fontenc}

\DeclareRobustCommand{\VAN}[3]{#2}
\let\VANthebibliography\thebibliography
\def\thebibliography{\DeclareRobustCommand{\VAN}[3]{##3}\VANthebibliography}

\usepackage{graphicx}	
\usepackage{amsmath}	
\usepackage{subcaption}
\usepackage{xspace}

\newcommand{\jwst}{\textit{JWST}\xspace}
\newcommand{\hst}{\textit{HST}\xspace}

\title[Globular Clusters in Abell 2744 from JWST]{\textit{JWST} Photometry of Globular Cluster Populations in Abell 2744 at $z=0.3$}

\author[W.~E.~Harris and M.~Reina-Campos]{
William E. Harris$^{1}$\thanks{E-mail: harris@physics.mcmaster.ca} and 
Marta Reina-Campos$^{1,2}$\thanks{E-mail:reinacampos@mcmaster.ca}
\\
$^{1}$Department of Physics \& Astronomy, McMaster University, 1280 Main Street West, Hamilton, L8S 4M1, Canada\\
$^{2}$Canadian Institute for Theoretical Astrophysics (CITA), University of Toronto, 60 St George St, Toronto, M5S 3H8, Canada\\
}

\date{Accepted XXX. Received YYY; in original form ZZZ}

\pubyear{2023}

\begin{document}
\label{firstpage}
\pagerange{\pageref{firstpage}--\pageref{lastpage}}
\maketitle

\begin{abstract}
\textit{JWST} imaging of the rich galaxy cluster Abell 2744 at $z=0.308$ has been used by the UNCOVER team to construct mosaic images in the NIRCAM filters. The exceptionally deep images in the ($F115W$, $F150W$, $F200W$) bands reveal a large population of unresolved pointlike sources across the field, the vast majority of which are globular clusters (GCs) or Ultra-Compact Dwarfs (UCDs). To the limits of our photometry, more than 10,000 such objects were measured, most of which are in the halos of the five largest A2744 galaxies but which also include GCs around some satellite galaxies and throughout the InterGalactic Medium  (IGM).  The measured luminosity function covers almost four magnitudes and follows a classic lognormal shape, though the GCLF turnover point still lies fainter than the photometric completeness limit. The colour index  ($F115W-F200W$) in particular covers an intrinsic spread of $0.5$ mag, clearly resolving the expected range of GC metallicities. The overall results are consistent with a large, normal GC population seen at a $3.5~$Gyr earlier stage of dynamical  evolution. 
\end{abstract}




\section{Introduction}



Abell 2744 (Pandora's Cluster), an extremely rich cluster of galaxies at redshift $z=0.308$, has been the target for intensive imaging campaigns with \hst (\textit{Hubble Space Telescope}) and now more recently with \jwst (\textit{James Webb Space Telescope}). As a massive lensing cluster and one of the six Hubble Frontier Fields \citep{lotz+2017}, it is an attractive testbed for detection and deep imaging of high-redshift galaxies and for gravitational lensing phenomena in general.  

The UNCOVER project team \citep{bezanson22} has made publicly available a set of deep mosaic images of A2744 generated from the \jwst NIRCAM, \hst ACS, and \hst WFC3 cameras. The central part of this field is shown in Figure \ref{fig:field}.  This material has been used to extract a catalog of 50,000 galaxies in the region \citep{weaver+2023}.  Close inspection of the NIRCAM SWC (Short Wavelength Channel) mosaic images also shows the presence of rich populations of unresolved point sources especially concentrated around the major galaxies in the cluster.  These are the globular clusters (GCs) that are found around every luminous galaxy \citep[e.g.][]{harris2015}, along with some probable Ultra-Compact Dwarfs (UCDs).  A small portion of the field is shown in Figure \ref{fig:sample} to illustrate the presence of these objects.   

A2744 is one of the most well studied clusters of galaxies at intermediate redshift, having attracted extensive observations across all wavelength regimes as well as weak and strong lensing analyses.  There are five major BCG-like galaxies in the field, two to the southeast (usually adopted as the cluster core in previous studies), another pair to the west, and a single one to the north (Fig.~\ref{fig:field}).  The position and velocity distributions of the member galaxies show considerable substructure, indicative of highly active merging in progress \citep[e.g.][]{owers+2011,jauzac+2016,medezinski+2016}.  A large X-ray and radio halo is somewhat more symmetric, with a single major center close to the core BCG pair, but has substructure as well \citep{kempner_david2004,boschin+2006,owers+2011,eckert+2015,jauzac+2016}.  The overall interpretation is that the subclusters have already experienced at least one passage, with a final merger into a single virialized system still very much in progress, though the details of the merging geometry and history are not yet settled \citep{merten+2011,owers+2011,eckert+2015,kimmig+2023}. 

Accounting for the substructure and globally non-virialized nature of the system, the total mass of A2744 is determined from recent studies to be $M_{200} \simeq 2 \times 10^{15} M_{\odot}$ \citep{boschin+2006,jauzac+2016,medezinski+2016}.  The corresponding virial radius is $R_{200}$ = 2.6 Mpc.  By comparison with the two most well known nearby clusters of galaxies, Virgo has a mass $5.5 \times 10^{14} M_{\odot}$ \citep{durrell+2014}, while Coma is at $\simeq 1.3 \times 10^{15} M_{\odot}$ \citep[e.g.][]{hughes1989,lokas_mamon2003,kubo+2007,ho+2022}.

At the redshift of A2744 the lookback time is $3.5$ Gyr, a quarter of the way back to the very first stages of galaxy formation.  The exceptionally deep images from \jwst give us an opportunity to observe very directly the characteristics of entire GC systems at an earlier stage in their evolution.  The ability to study large populations of GCs like this is entirely complementary to recent \jwst observations of objects like the Sparkler, a lensed galaxy at $z=1.38$ in which a handful of \emph{individual} GCs or proto-GCs at even earlier stages of evolution have been seen \citep{mowla+2022,claeyssens+2023,forbes_romanowsky2023,adamo+2023}; see also \citet{vanzella+2017,vanzella+2019,vanzella+2022} for examples in other systems. 

The \textit{Hubble Space Telescope} (\hst) has been a powerful instrument for surveying GC populations around galaxies out to limits of roughly twice the Coma cluster distance, i.e.~$\lesssim 200$ Mpc or $z \lesssim 0.05$ \citep[e.g.][]{peng+2011,lim+2018,madrid+2018,amorisco+2018,harris2023,hartman+2023,dornan_harris2023}.  For all these cases, cosmological lookback times are $\lesssim 0.6$ Gyr and we are essentially observing within just the Local Universe. In a few rare cases with very long exposures, \hst photometry has been obtained for GCs  and UCDs in more distant systems including an unnamed elliptical galaxy at $z=0.09$ \citep{kalirai+2008}, the rich cluster A1689 at $z=0.18$ \citep{mieske+2004,alamo-martinez13,alamo-martinez+2017}, and A2744 itself \citep{blakeslee+2015,lee_jang2016,janssens+2017}.  
Even so, in these cases only the very brightest $\sim 1$ magnitude or so of the GC population is recovered with any confidence. The first examples of JWST photometry for entire populations of GCs/UCDs in distant environments are by \citet{faisst+2022} and \citet{lee22} for the cluster SMACS J0723.3-7327 at $z=0.39$.  However, the UNCOVER database for A2744 reaches much deeper in absolute magnitude into the star cluster population.

An especially interesting new question is the link between the spatial distribution of GCs and the dark matter (DM) distribution of the galaxy halos and the cluster they are embedded within. Recent theoretical studies \citep[e.g.][]{alonso+2020,reina-campos+2022,reina-campos+2023} reinforce the view that the GC distributions and the IntraCluster Light can be used to recover the DM profile. Observational tests of such links include the large-scale survey of the Virgo cluster \citep{durrell+2014} and \jwst imaging of the cluster SMACS J0723.3-7327 at $z=0.39$ \citep{lee22,diego23}. Carrying out this kind of comparison is 
opened up by the ability to measure large-scale spatial distributions of GCs in clusters at significant redshifts, an observational parameter space well suited to \jwst.

In the present study, we describe photometry of the A2744 NIRCAM SWC images directed specifically at isolating the pointlike sources in the field, which are dominated by the GCs around the member galaxies. Here, we present the measurements and give a brief overview of the system. In following studies, we will analyze the GC spatial distribution and their distributions in metallicity and luminosity more quantitatively. Section \ref{sec:data} describes the data and the photometric measurement steps; Section \ref{sec:colour-distribution} presents the GC colour distribution function and a brief analysis; and Section \ref{sec:luminosity-function} presents the luminosity distribution. Section \ref{sec:adjustments-z0} gives a brief discussion of the adjustments necessary to normalize the data to zero redshift.  Section \ref{sec:discussion} discusses connections of our data to a potential UCD population.  Finally, Section \ref{sec:summary} gives a summary of our findings and prospects for additional work.

We adopt cosmological parameters $H_0 = 67.8$ km s$^{-1}$ and $\Omega_{\Lambda} = 0.692$, which for $z = 0.308$ gives a luminosity distance $d_L = 1630$ Mpc or $(m-M)_0 = 41.06$ for A2744, and an angular size distance of $d_A = d_L/(1+z)^2 = 952$ Mpc. The foreground extinction (from the NED database) is $A_V = 0.036$, which has a negligible effect for the NIR bandpasses used in the present study.  

\section{Data and Photometry}\label{sec:data}

\begin{figure}
  \centering
  \begin{subfigure}[t]{0.38\textwidth}
  \vspace{0pt}
  \hspace{15pt}
  \includegraphics[width=0.92\linewidth, trim=1.0cm 0 0 0]{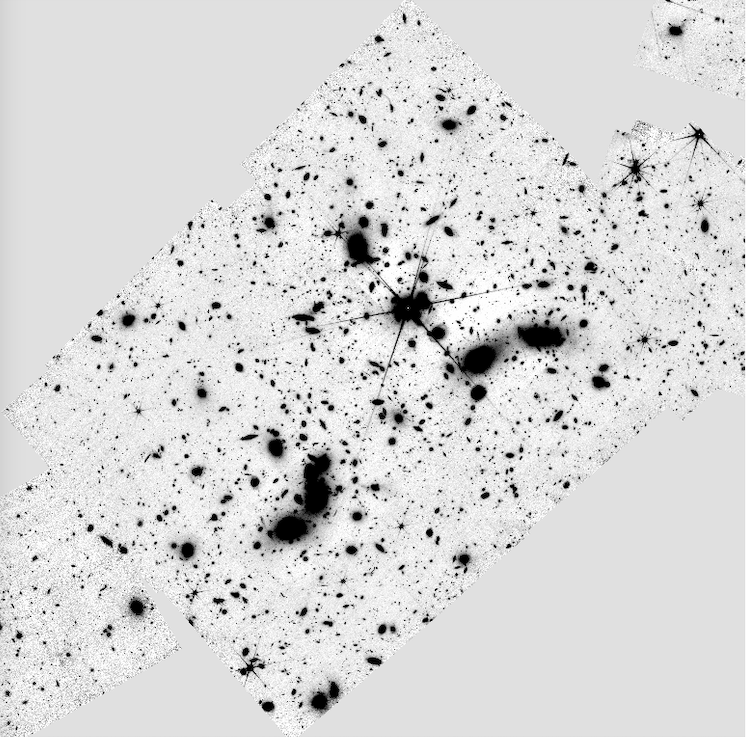}
  \end{subfigure}
  \begin{subfigure}[t]{0.49\textwidth}
  \vspace{-60pt}
  \includegraphics[width=0.95\linewidth]{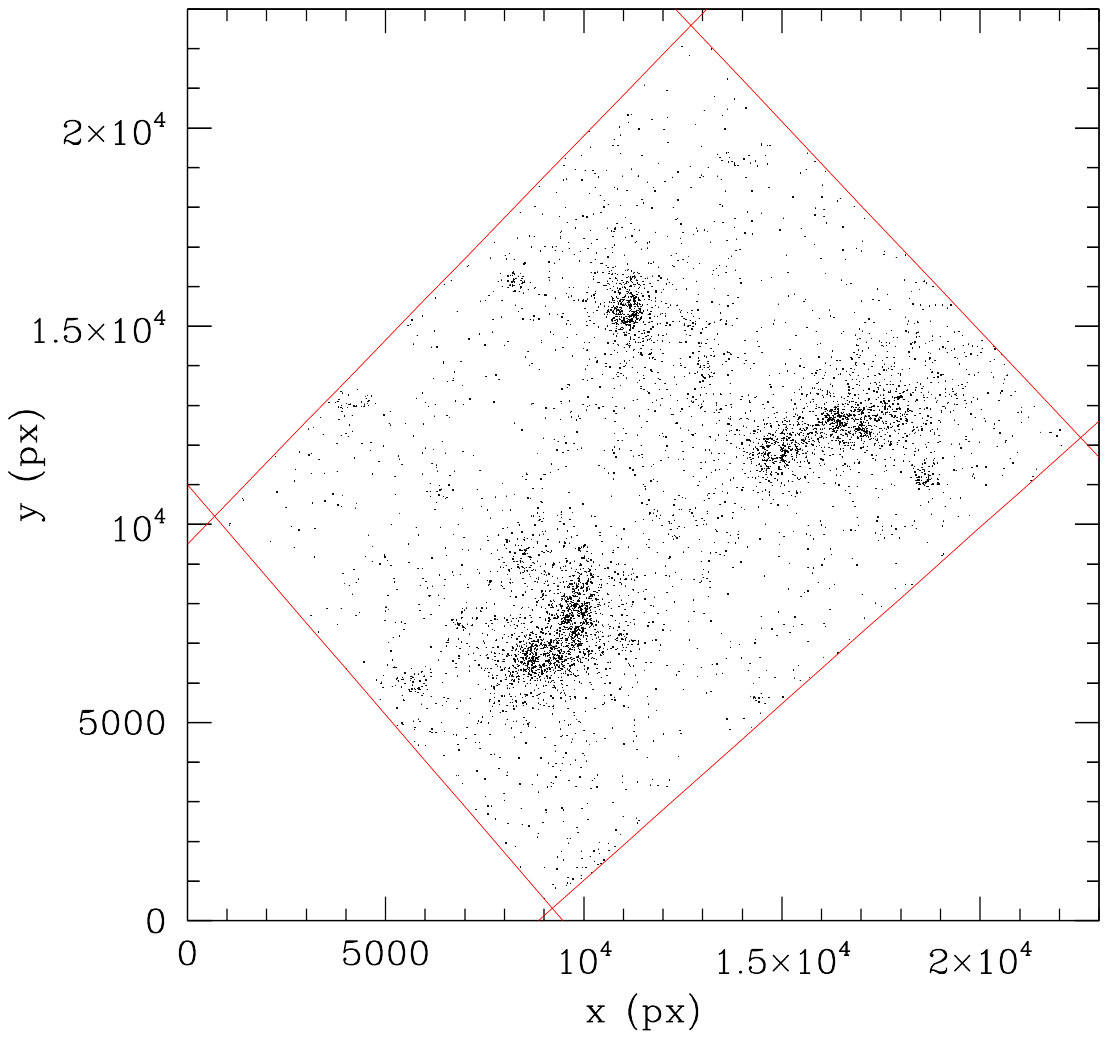}
  \end{subfigure}
  \vspace{-2.2cm}
  \caption{\emph{Upper panel:}  The central part of the UNCOVER mosaic field of Abell 2744, with North at top and East at left.   \emph{Lower panel:} Spatial distribution of the candidate globular clusters measured in this study. The red quadrangle marks the boundary for the central region discussed in the text. As described in the text, parts of the mosaic image to the lower left and upper right of the box with shorter total exposures and higher sky noise were not used.  The region shown is $460'' = 2100$ kpc across.}\label{fig:field}
\end{figure}

\begin{figure*}
	\includegraphics[width=0.95\textwidth]{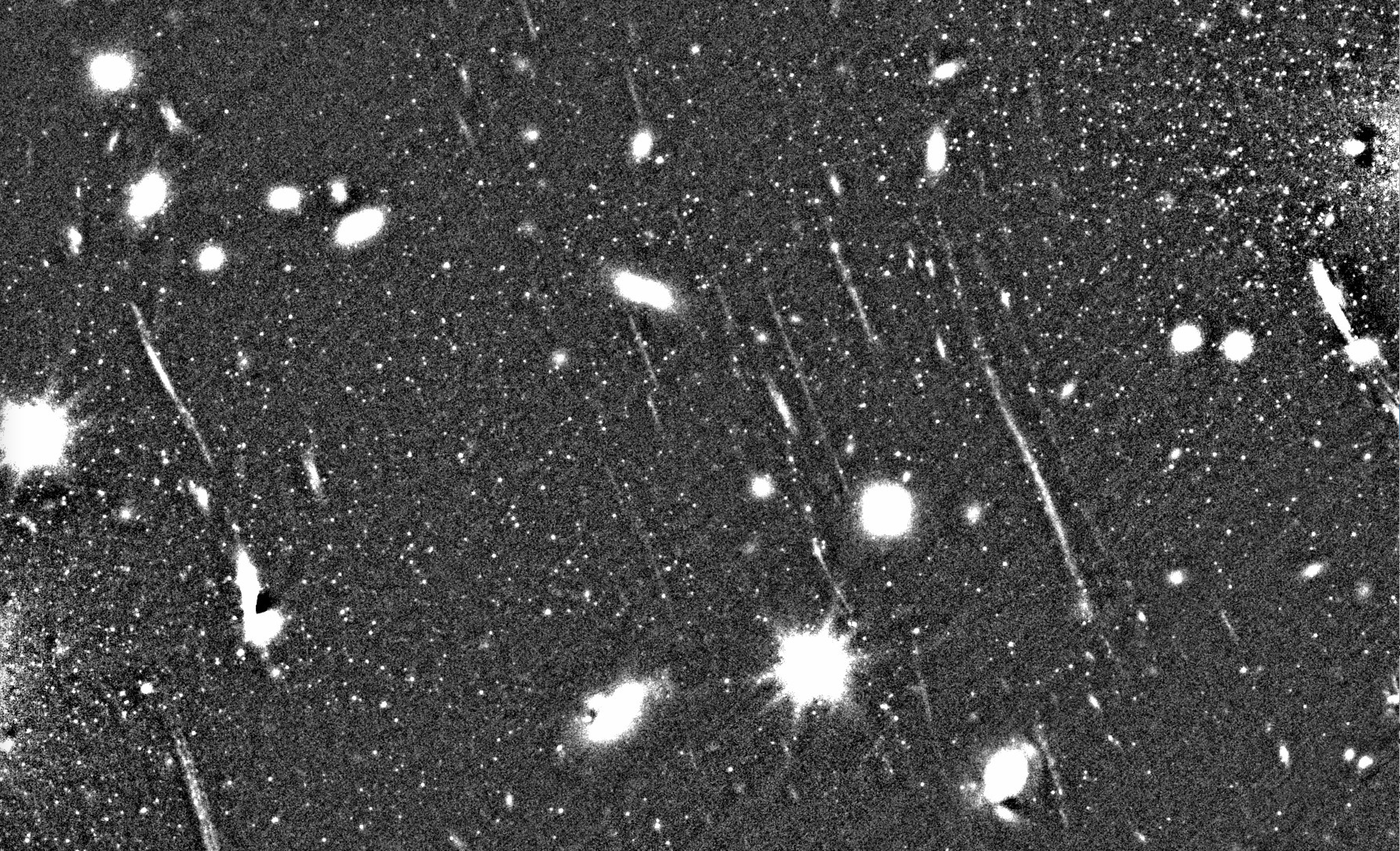}
  \caption{A small portion of the A2744 mosaic field shown at very high contrast, at a location between the two major galaxies at middle right of Figure 1.  The field shown is $32''$ across, equivalent to 150 kpc linear distance.  The image has been median filtered to remove the large-scale light gradients from the galaxies and to emphasize the point sources.  The many small starlike objects indicate some of the globular cluster and UCD candidates. The thin short diagonal arcs are strongly lensed background galaxies.} \label{fig:sample}
\end{figure*}

The mosaic images of the A2744 field and their construction are thoroughly described in \citet{bezanson22}.  We obtained the NIRCAM SWC images in $F115W$, $F150W$, and $F200W$, all three of which had closely similar depths (limiting magnitudes) for objects like GCs with intermediate colours. The $F090W$ image does not cover as large an area and has shallower depth, and was not used here. Similarly, the NIRCAM images from the Long Wavelength Channel (LWC), with their larger pixel scale and lower resolution, also did not reach sufficient limiting magnitudes for our purposes and were not used. The mosaics, built from NIRCAM exposures in three different programs (GO-2561, ERS-1324, DD-2767), have an irregular overall shape with different total exposure times in various regions \citep[see particularly Fig.~2 of][]{bezanson22}, but the central region thoroughly covering the cluster core is included by all the programs. Our analysis will concentrate on that central region, which is marked out in Fig.~\ref{fig:field}b.  The other, outlying wings of the mosaics have larger levels of sky noise and inhomogeneity and were not used \citep[see][]{bezanson22}.

The mosaic images in $F115W$, $F150W$, and $F150W$ have a resampled pixel size of $0.02''$, smaller than the native SWC camera scale of $0.03078''$ per pixel. The angular scale is therefore 92 pc per $0.02''$ pixel. The mosaics are reoriented to align with the cardinal directions with EW along the x-axis and NS along the y-axis. Direct measurements of unsaturated stars in the images give PSF (point spread function) sizes of \emph{fwhm} $= 1.9$ px ($F115W$), 2.4 px ($F150W$), and 3.0 px ($F200W$). 

Photometry was carried out with the tools in \texttt{daophot} \citep{stetson1987} in their IRAF implementation.  The first step was to add the images in the three filters to produce a very deep stacked image for object finding. On this coadded image, we used the \emph{daofind} package to generate a finding list of about $10^6$ candidate objects.  Aperture photometry with \emph{phot} and a 3-px ($0.06''$) radius was then carried out with that list on each filter.

The next stage of the photometry is PSF fitting to the object list. Normally, bright but unsaturated foreground stars can be used for constructing a PSF, but they are rare in this high-latitude field. However, the GC/UCD population in A2744 itself provides a sample of unresolved (starlike) objects; they are numerous, and the brightest are more than three magnitudes above the photometric limits. At the distance of A2744, a typical GC with a half-light diameter $2 r_h \sim 5$ pc has an angular diameter of $0.0011'' =$ 0.054 px on these images, almost two orders of magnitude below the resolution limit. Even UCDs with their scale sizes that can be up to $10 \times$ larger \citep[e.g.][]{drinkwater+2000,dabringhausen+2008} can be expected to be unresolved or at worst marginally resolved. About 100 such objects in the range $F150W$ $< 27.5$ scattered across the central region of the field were carefully selected and confirmed by radial profile measurement.  The \emph{pstsel} and \emph{psf} tools were then used to construct an empirical PSF from them for each filter. PSF fitting was then done on the complete photometry list with \emph{allstar}. 
  
Next, the lists of output photometry from the three filters were matched by position to within 2 px, accepting only objects that were detected in \emph{all three} filters. Though the three images have very similar depths, the triple-match requirement is useful for rejecting a large proportion of false detections including pixel artifacts, noise clumps, and the like. 

As a final step for reducing sample contamination, the \texttt{daophot} \emph{sharp} parameter was used to reject remaining objects that were not starlike (the rejected objects are mainly small, faint background galaxies, severely crowded objects, or pixel-scale artifacts that do not match the PSF profile) \citep[see][for additional discussion]{harris2009,harris2023}. 

Final measured magnitudes were calibrated to the ABMAG system following \citet{weaver+2023}. The fluxes per pixel on the mosaic images are in units of 10 nJy, from which the magnitudes are defined as
\begin{equation}
    m_{AB} = -2.5 \textrm{log}(f/\textrm{10 nJy}) + 28.90
\end{equation}
where $f$ is the total flux of the object.  We used the PSF encircled-energy tables as published on the NIRCAM webpages 
to help estimate the necessary aperture corrections from the the PSF-fitted instrumental magnitudes to large radius:  direct aperture photometry of the isolated PSF stars on the images was used to determine the step from the PSF-fitted magnitudes to a radius of $r=12.5$ px ($0.25''$) that encloses 85\% of the total profile light. Lastly, a further $ -0.18$ mag was added to step to `infinite' radius. 

After the photometry and culling steps, the list of GC/UCD candidates consists of 10617 objects. A final close visual inspection of these objects on the images showed that the list included virtually all starlike objects that are clearly detectable, while obviously nonstellar ones were almost all excluded. The final colour-magnitude diagrams (CMDs) for these matched, starlike sources appear in Figure \ref{fig:cmd}, where $F150W$ is plotted versus the three different colour indices that can be obtained from the combination of filters\footnote{The third panel of Fig.~\ref{fig:cmd} has the unusual property for a CMD that the x and y axes are independent of each other.}. Their spatial distribution across the central field of A2744 is shown in Figure \ref{fig:field}b.  The concentration of the objects around the five biggest galaxies is very evident, but smaller galaxies can be spotted as well, and many GC candidates are also scattered more widely through the ICM (IntraCluster Medium).

To gauge the effective photometric limits of the data and the internal measurement uncertainties, artificial-star tests were run. With the \texttt{daophot} \emph{addstar} tools, 16000 fake stars covering the magnitude range $26 - 34$ were inserted into the three images in series so that no change in the crowding levels on the images would result (though the absolute crowding level is quite low to begin with). The synthesized frames were remeasured with exactly the same procedure as described above.  The resulting recovery completeness is shown in Figure \ref{fig:f}, where the recovery fraction $f$ is defined as the number of artificial stars measured in a given $0.2-$mag bin, divided by the number of stars originally inserted into that bin.  Defined this way, the ratio $f$ implicitly includes the effects of bin-jumping; i.e.~a measured star falling within the bin may have originally been assigned to another adjacent bin \citep{alamo-martinez13}.  Note as well that in Fig.~\ref{fig:f}, $f$ is plotted versus the measured (i.e., recovered) magnitude. For $f > 0.2$, the trend is accurately matched by a modified hyperbolic tangent function 
\begin{equation}
  f(m) = \frac{2}{1+\exp\left[{\beta (m-m_1)}\right]} - 1  
  \label{eq:f}
\end{equation}
with $\beta = 1.70, m_1 = 30.28$.\footnote{This function has the unphysical property that $f$ becomes negative for $m > m_1$, but its most important role is to match the bright end and the downturn region.  No part of our analysis depends on data fainter than F150W $\simeq 30$.}  The $50~$per cent recovery completeness point is at $F150W$ = 29.63, and $80~$per cent at $F150W$ = 28.98.  The artificial-star tests were also used to gauge the trend of photometric measurement uncertainty $\sigma$ with magnitude, as shown in the second panel of Fig.~\ref{fig:f}.  A simple exponential curve $\sigma(m) \simeq 0.075 \exp\left[{0.7(m-29.0)}\right]$ describes the trend well in each filter, to the useful limits of the data.

\section{Colour Distributions}\label{sec:colour-distribution}

\begin{figure*}
	\includegraphics[width=0.70\textwidth]{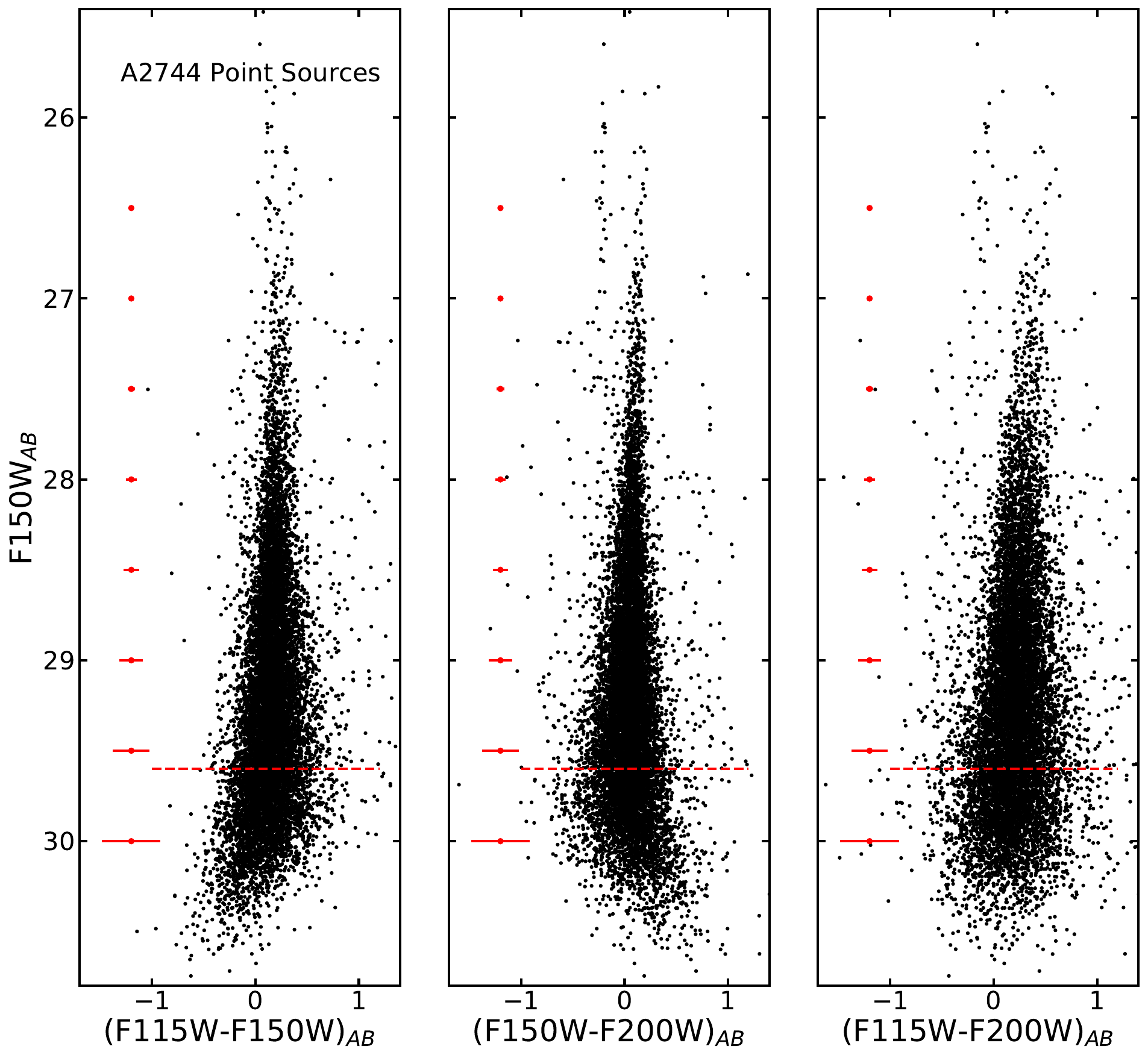}
    \caption{CMDs for the unresolved objects in the A2744 field after triple-matching among the three filters and selection by \emph{allstar/sharp} to remove nonstellar objects. This culled sample is dominated by GCs around the A2744 galaxies. Errorbars along the left side of each diagram show the expected $\pm 1\sigma$ spread due to measurement uncertainty. The $50~$per cent recovery completeness level at $F150W = 29.6$ is shown as the dashed line.} \label{fig:cmd}
\end{figure*}

\begin{figure}
	\includegraphics[width=0.80\columnwidth]{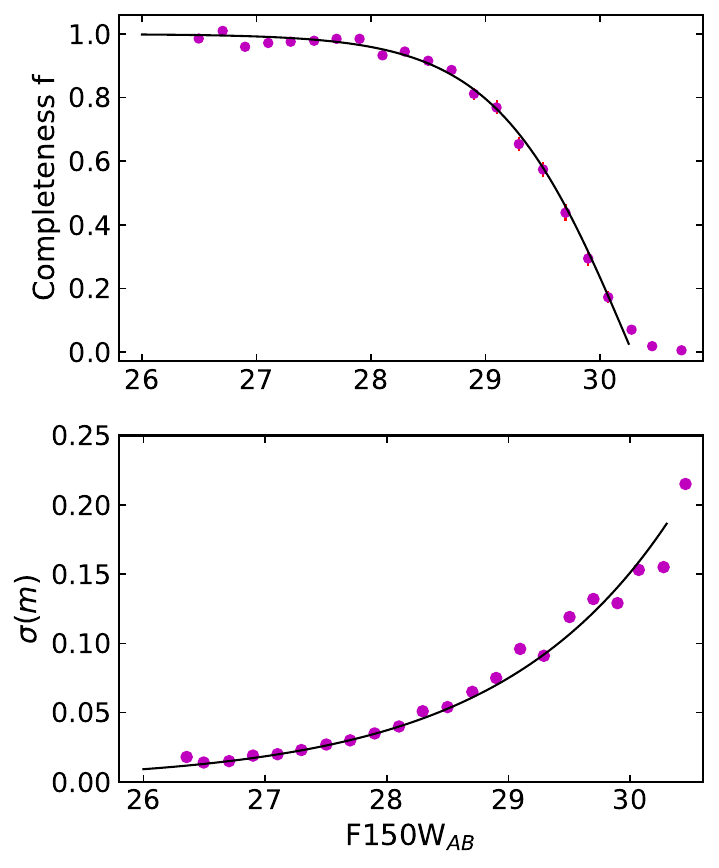}
    \caption{\emph{Upper panel:} Photometric recovery completeness, measured from the artificial-star tests. The recovery fraction $f$ is plotted in $0.2-$mag bins. The solid line shows the tanh function in Eq.~\ref{eq:f} with parameters $\beta=1.70$, $m_1=30.28$.
    \emph{Lower panel:} Internal measurement uncertainty $\sigma$ versus magnitude, as determined from the artificial-star tests. The exponential curve given in the text is shown as the solid line.} \label{fig:f}
\end{figure}

The integrated colours of GCs in the NIRCAM filters are relatively new in this subject, with little available data \citep{faisst+2022,lee22}. By comparison, there are available observations in many nearby galaxies of \hst photometry of GCs in several ACS and WFC3 filters \citep[e.g.][]{harris2023,hartman+2023}.
GC colours depend primarily on the cluster \emph{metallicity} and \emph{age}; increases in both quantities drive the cluster to redder colours. However, as will be seen below, the NIRCAM colour indices in adjacent filters such as $(F150W-F200W)$ and $(F115W-F150W)$ are not strongly sensitive to metallicity, so for the CMDs in these colours we should expect to find that old GCs delineate a narrow, vertical sequence exactly like what is seen in Fig.~\ref{fig:cmd}a-b.  The colour $(F115W-F200W)$, however, is about twice as metallicity-sensitive as either of these and the GC sequence is distinctly broader (Fig.~\ref{fig:cmd}c).

\begin{figure}
	\includegraphics[width=\columnwidth]{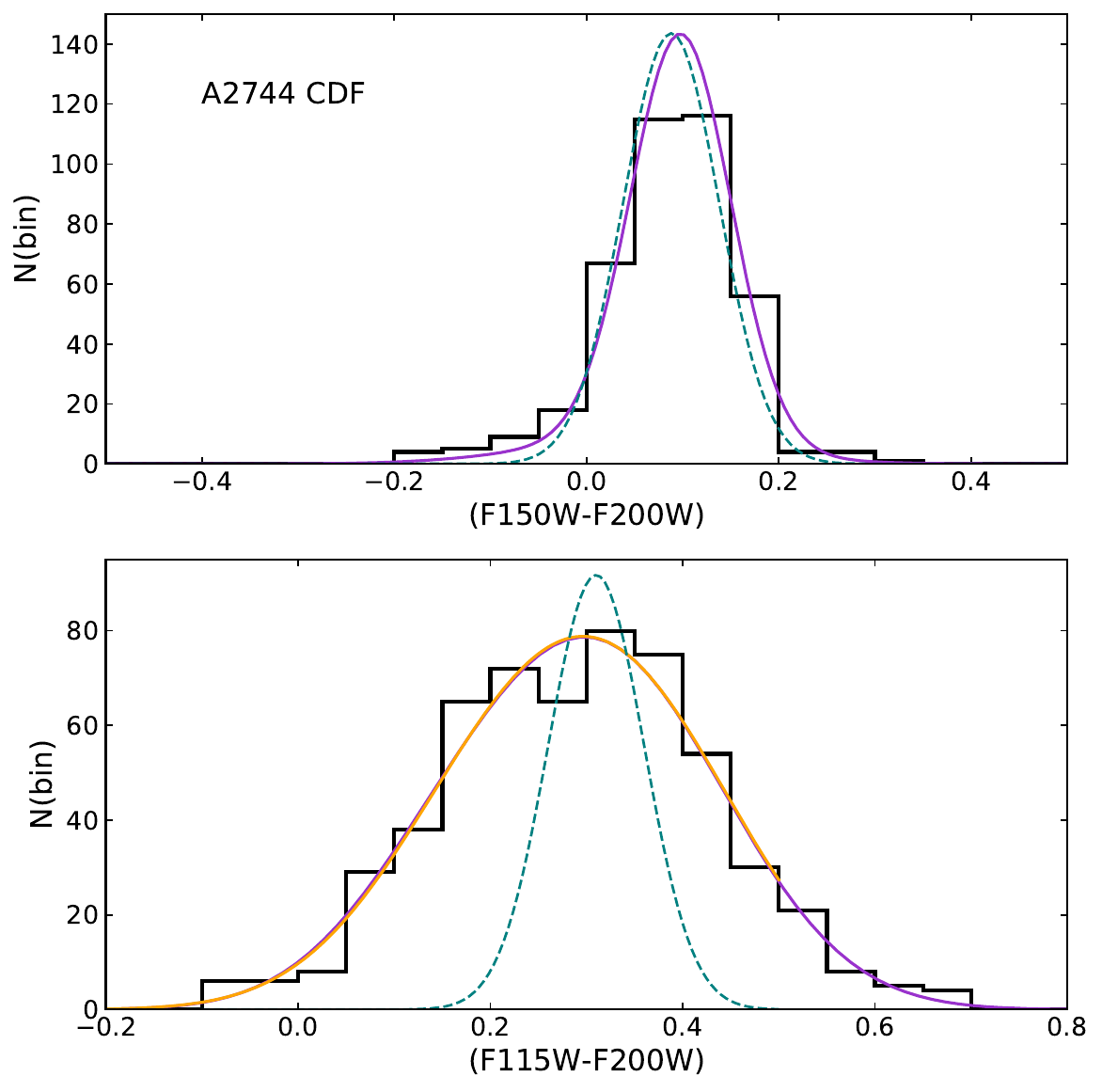}
    \caption{\emph{Upper panel:} Histogram of the GC colours in ($F150W$--$F200W$), for objects in the magnitude range $F150W = 27.0 - 28.5$ for which completeness is $>80~$per cent.  The solid line shows the double-Gaussian fit described in the text. The green dashed-line Gaussian centered on the mean colour shows the expected maximum dispersion if the colour spread was dominated purely by measurement scatter.  \emph{Lower panel:} Histogram of GC colours for the same sample in ($F115W-F200W$). The solid line shows the double-Gaussian fit to the data (either homoscedastic or heteroscedastic).  NB: K-corrections have \emph{not} been applied to the colour indices; see text. } \label{fig:cdf}
\end{figure}

A basic test of the observed range of colours is shown in Figure \ref{fig:cdf}, displaying the histograms over the magnitude range $27.0 < F150W < 28.5$ where the photometry is the most precise. In this range the recovery completeness is higher than 80\%. From many previous studies drawn primarily from optical colour indices, we expect the GCs to fall into two identifiable metallicity groups, `blue' (metal-poor) and `red' (metal-rich) with a double-Gaussian model usually providing an accurate match to the colour distribution function (CDF) \citep[e.g.][among many others]{larsen+2001,peng+2006,harris2009,faifer+2011,brodie+2012,cho+2012,hargis_rhode2014,harris2023,hartman+2023}.  For either the index $(F150W-F200W)$ or $(F115W-F150W)$, the observed total colour range is no more than $\sim 0.2$ mag and is not much larger than what would be produced just by random measurement scatter. In Fig.~\ref{fig:cdf}a, the dashed line shows the shape of the CDF expected if the intrinsic colours of the GCs were dispersionless and the broadening was equal to the $1 \sigma$ colour uncertainty at $F150W=28.0$, i.e.~the \emph{maximum} colour dispersion over the given magnitude range. For comparison, the solid line gives a double-Gaussian fit to the CDF determined from the Gaussian mixture model (GMM) fitting code \citep{muratov_gnedin2010}; it is scarcely different from the single-Gaussian error distribution, so there is no compelling evidence for more than one component.

For the broader colour index $(F115W-F200W)$, however (Fig.~\ref{fig:cdf}b), the total width of the CDF is $\sim 0.5$ mag and the rms dispersion is $\pm0.15$ mag, larger than the $\pm0.05-$mag rms spread expected from the measurement uncertainties alone (dashed line in the Figure). The best-fit double Gaussian solution performed with GMM gives the results listed in Table \ref{tab:gmm}, where $\mu_1, \mu_2$ are the means of the bluer and redder subcomponents; $\sigma_1, \sigma_2$ are their dispersions; $f_2$ is the fraction in the redder component;
and $p(uni)$ is the probability that the CDF is a single Gaussian.
Their sum is shown as the solid (magenta) line in Fig.~\ref{fig:cdf}b.  Clearly, a single-Gaussian solution is also acceptable.  A simpler homoscedastic fit (enforcing equal variances for the two components) gives ($\mu_1 = 0.244, \mu_2 = 0.393$) and ($\sigma_1 = \sigma_2 = 0.098$), with $f_2 = 0.44$. The quality of fit between the two solutions is nearly identical. The main conclusions drawn from this exercise are that (a) the intrinsic width of the CDF is resolved by the ($F115W-F200W$) index in particular; i.e.~its spread is distinctly larger than can be accounted for by measurement scatter alone; and (b) either a single-Gaussian or a conventional double-Gaussian model can provide an accurate match to the CDF, though the blue/red fraction in particular is sensitive to the assumptions made about the parameters.

\begin{figure}
	\includegraphics[width=\columnwidth]{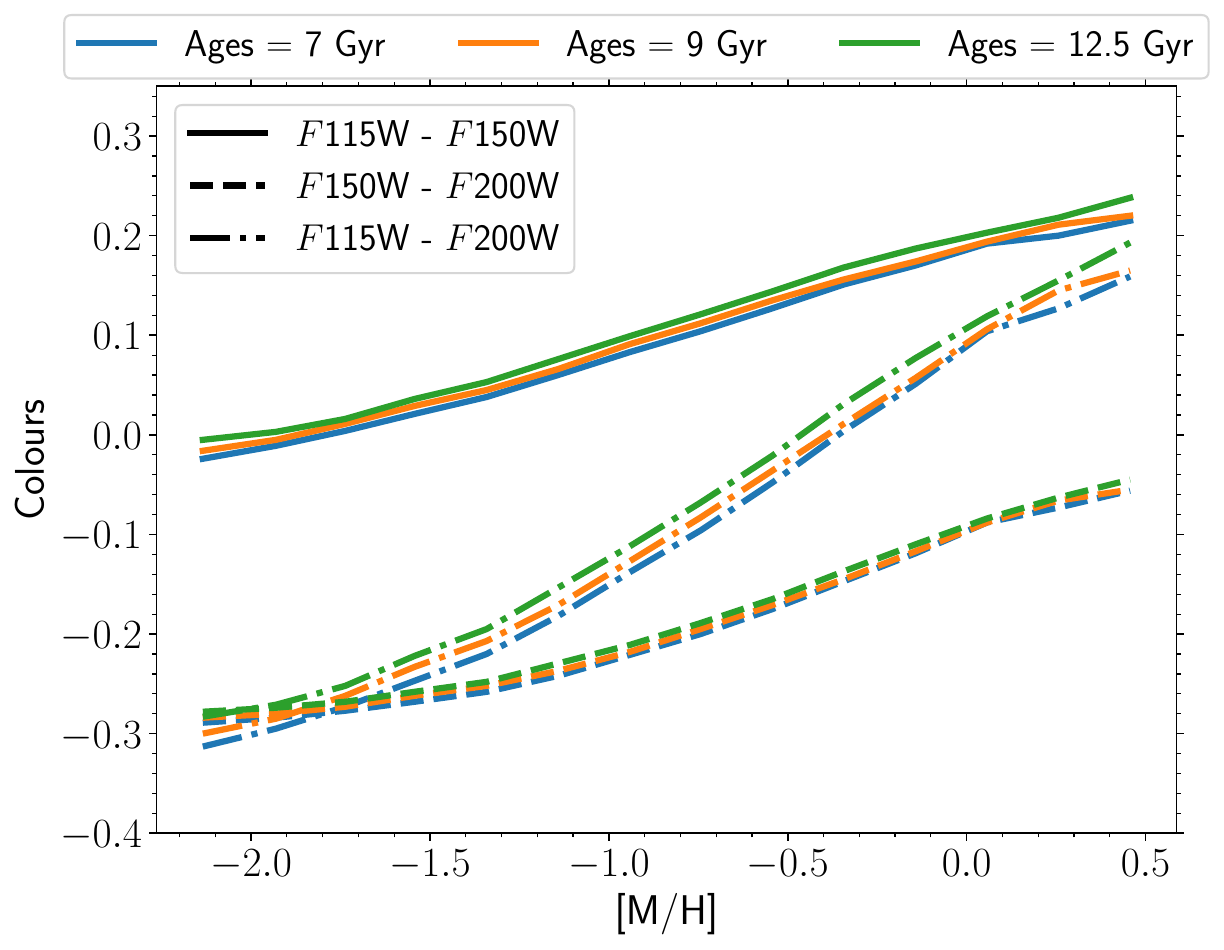}
    \caption{Predicted colour-metallicity relations for globular clusters in the NIRCAM filters, from the PARSEC CMD3.7 stellar models. Green lines are for an assumed age of 12.5 Gyr, while orange lines are for 9.0 Gyr and blue lines for 7.0 Gyr. Different linestyles indicate different filter combinations as noted in the legend. Indices are in the ABMAG system.} \label{fig:color}
\end{figure}

In Figure \ref{fig:color}, theoretically predicted trends of colour versus metallicity are shown for the three indices used here. These tracks have been generated with the PARSECv1.2 CMD3.7 stellar models\footnote{http://stev.oapd.inaf.it/cgi-bin/cmd} \citep{bressan+2012} for three different ages (12.5 Gyr, 9.0 Gyr, 7.0 Gyr), and adopting the standard \citet{kroupa2001} IMF. Conversions of these models from their normal output Vegamag magnitude scale to the ABMAG scale have been applied\footnote{https://jwst-crds.stsci.edu}. For star clusters in this age range, colour is insensitive to age, even more so than for optical colour indices.

The best-fit single Gaussians to each of the CDFs give $\mu=0.088, \sigma=0.072$ for $(F150W-F200W)$, and $\mu = 0.290, \sigma=0.14$ for $(F115W-F200W)$.  After K-corrections to the colours (see \S\ref{subsec:k-correction} below), these mean colours adjusted to zero redshift become $\mu_0(F150W-F200W) = -0.16$ and $\mu_0(F115W-F200W) = +0.04$.  In both cases, comparison with Fig.~\ref{fig:color} would give mean metallicities near [M/H] $\sim -0.4$, though the comparison should be viewed with caution since it depends on all three of the K-correction accuracies, the absolute accuracy of the photometry including the aperture corrections, and the zeropoints of the model colour scales themselves.  The \emph{range} of colours may be a more robust comparison with the models.  These ranges are equal to $\Delta(F150W-F200W) \simeq 0.2$ mag and $\Delta(F115W-F200W) \simeq 0.5$ mag (Fig.~ \ref{fig:cdf}).  These correspond to an end-to-end range of 2.0 dex in metallicity, as observed for the GCs in most large galaxies \citep{peng+2006,harris2023}.   

\begin{table*}
\centering
\caption{GMM Fits to the Colour Distributions: $N$ is the number of objects, $\mu_i$ and $\sigma_i$ are the mean colour and its dispersion for the blue ($i = 1$) and red ($i=2$) components, $f_2$ is the fraction in the redder component, and $p(uni)$ is the probability that the CDF is a single Gaussian. Internal $1\sigma$ fitting uncertainties are given in parentheses.} 
\begin{tabular}{lccccccc}
  \hline \hline
Region & $N$ & $\mu_1$ & $\sigma_1$ & $\mu_2$ & $\sigma_2$ & $f_2$ & $p(uni)$ \\
   \hline
All & 566 & 0.152 (0.084) & 0.106 (0.031) & 0.320 (0.077) & 0.128 (0.030) & 0.82 (0.24) & 0.95 \\
5 BCGs & 413 & 0.166 (0.044) & 0.068 (0.021) & 0.341 (0.071) & 0.112 (0.018) & 0.84 (0.22) & 0.35 \\
IGM & 153 & 0.171 (0.050) & 0.121 (0.029) & 0.353 (0.157) & 0.156 (0.055) & 0.33 (0.28) & 0.29 \\
                 \hline
\end{tabular}
\label{tab:gmm}
\end{table*}

One additional comparison can be made directly with other \jwst photometry: \citet{faisst+2022} and \citet{lee22} have measured the brightest $\sim 1$ mag of the GC population in the $z=0.39$ cluster SMACS J0723.3-7327.  The mean colour (dereddened, but not K-corrected) in $(F150W-F200W)_{0,AB}$ for their sample is in the range $\simeq 0.05-0.15$ depending on location, in close agreement with our results for the A2744 population. The brightest GCs in their data are near $F200W_0 \simeq 27$, again very consistent with our A2744 CMD given that SMACSJ0723 is 0.6 mag more distant.

\begin{figure}
	\includegraphics[width=\columnwidth]{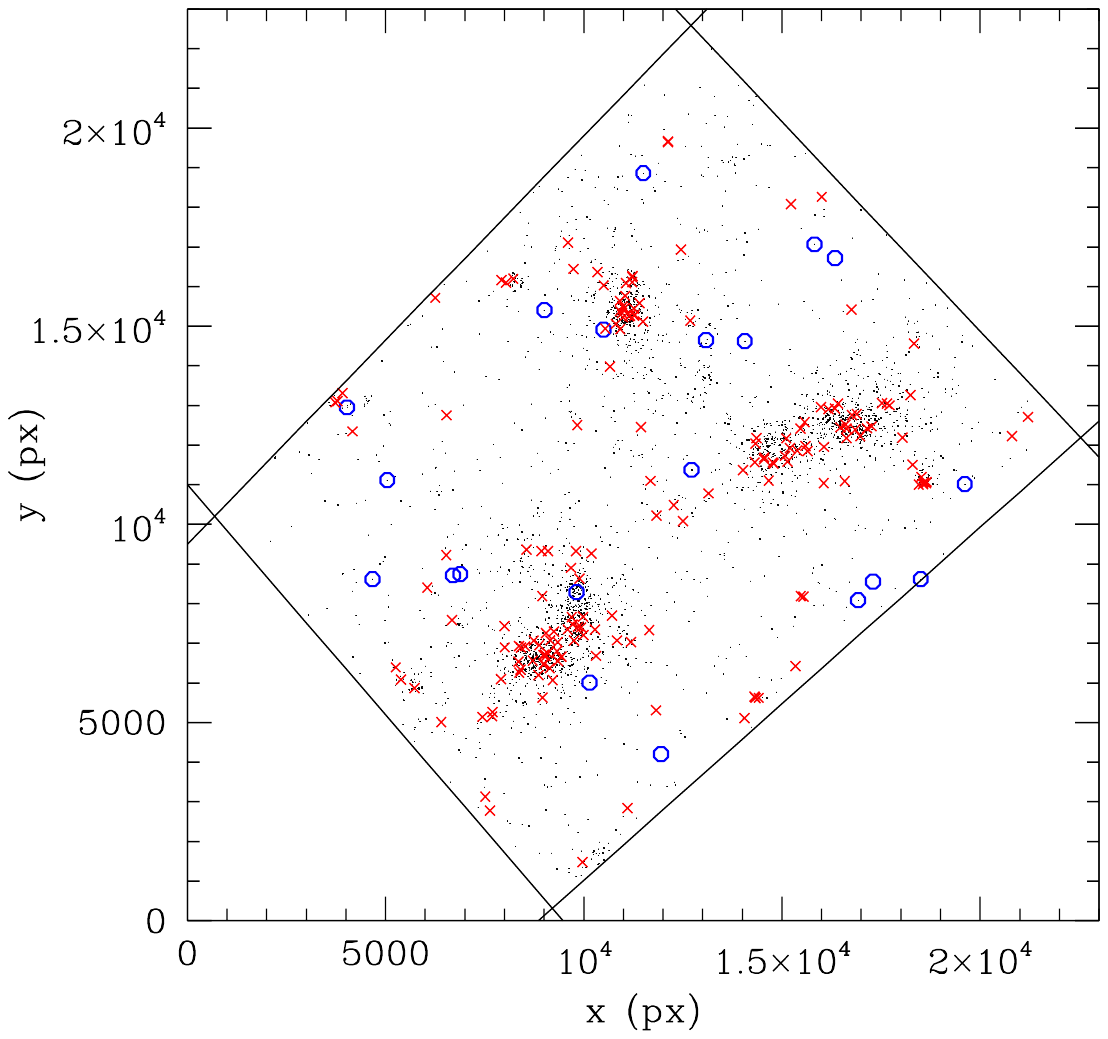}
 \vspace{-90pt}
    \caption{Locations of objects in the `blue sequence' described in the text are shown as blue circles, while the GC candidates brighter than $F150W = 29.0$ are small black dots.  The most luminous GC candidates at $F150W < 27.5$ are marked as red crosses.} \label{fig:xyblue}
\end{figure}

\begin{figure*}
	\includegraphics[width=0.90\textwidth]{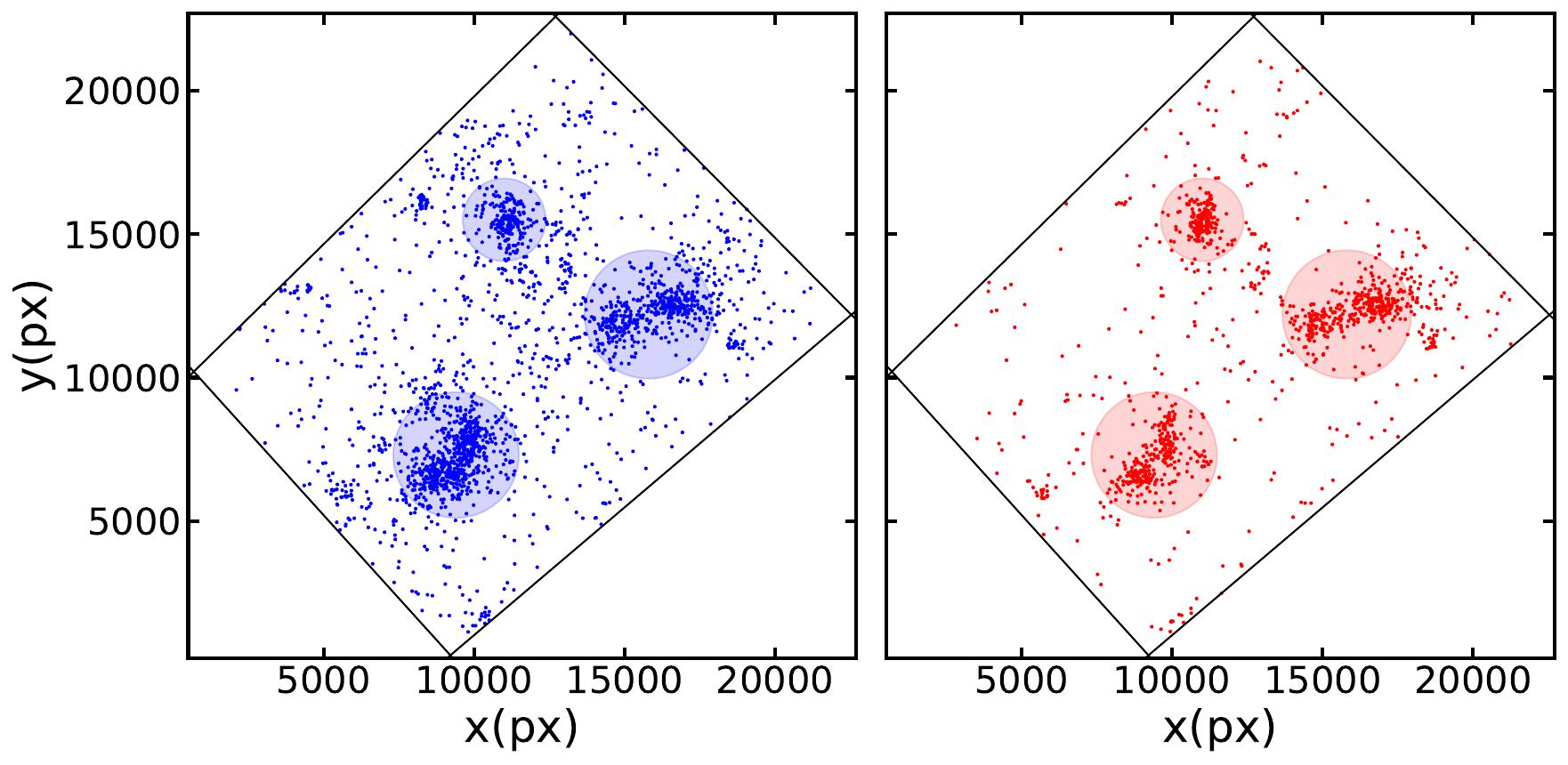}
    \caption{Spatial distributions of GC and UCD candidates with $F150W <29.0$. \emph{Left panel:} `Blue' GCs: those with colours $(F115W-F200W) < 0.31$. \emph{Right panel:} `Red' GCs: those with colours $(F115W-F200W) > 0.31$.  The circles mark out the exclusion regions around the five major galaxies as discussed in the text.} \label{fig:xypair}
\end{figure*}

\begin{figure}
\hspace{10pt}
  \includegraphics[width=0.95\columnwidth, trim=1.0cm 0 0 0]{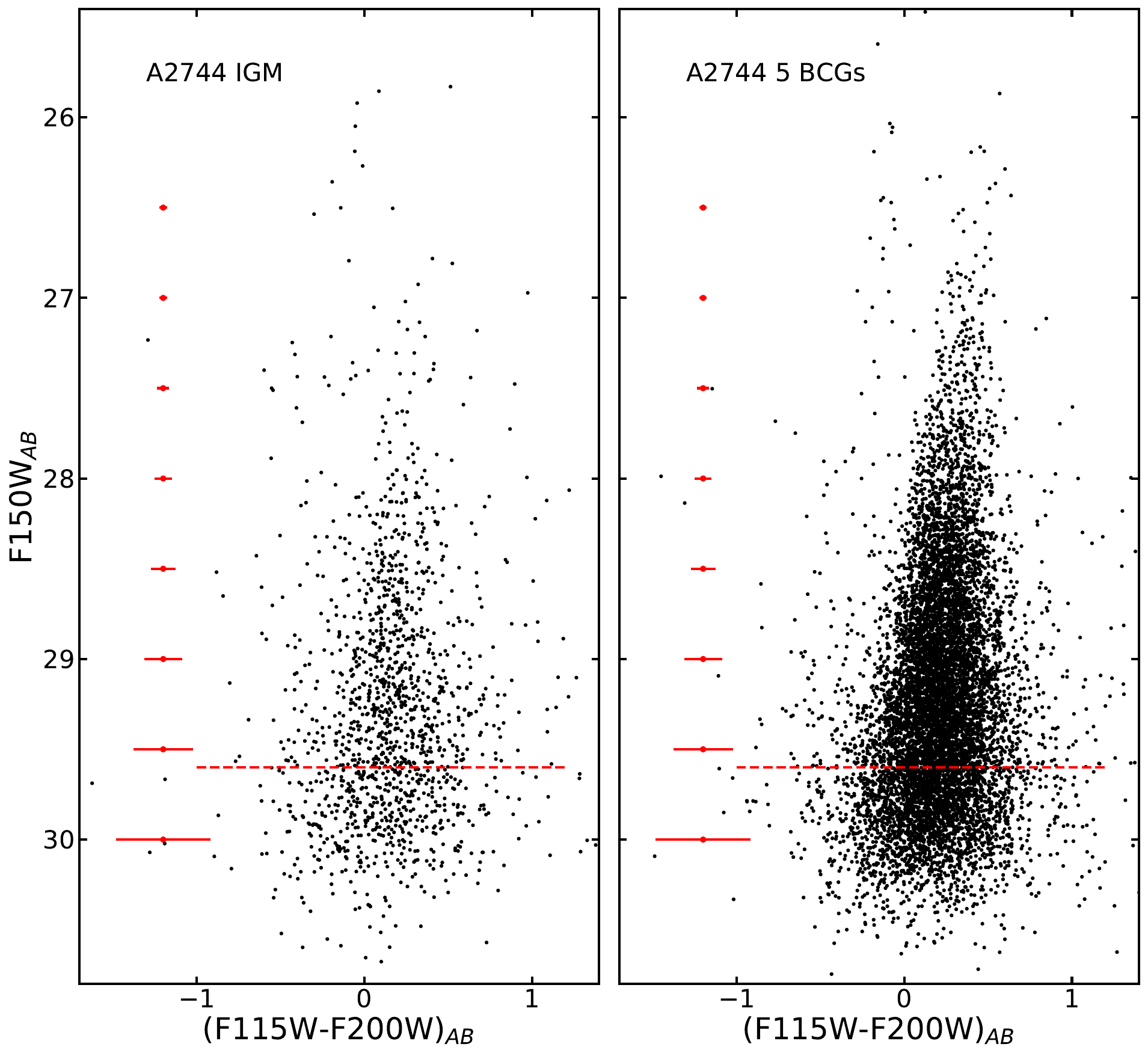}
  \caption{\emph{Left panel:} CMD for the objects outside the exclusion circles marked in Fig.~\ref{fig:xypair}.  This sample is dominated by GCs in the InterGalactic Medium and in a few small satellite galaxies. \emph{Right panel:}  CMD for the objects inside the circles.  These are dominated by members of the five giant galaxies in the system.  } \label{fig:cmd_igm}
\end{figure}

\begin{figure}
  \includegraphics[width=0.92\columnwidth]{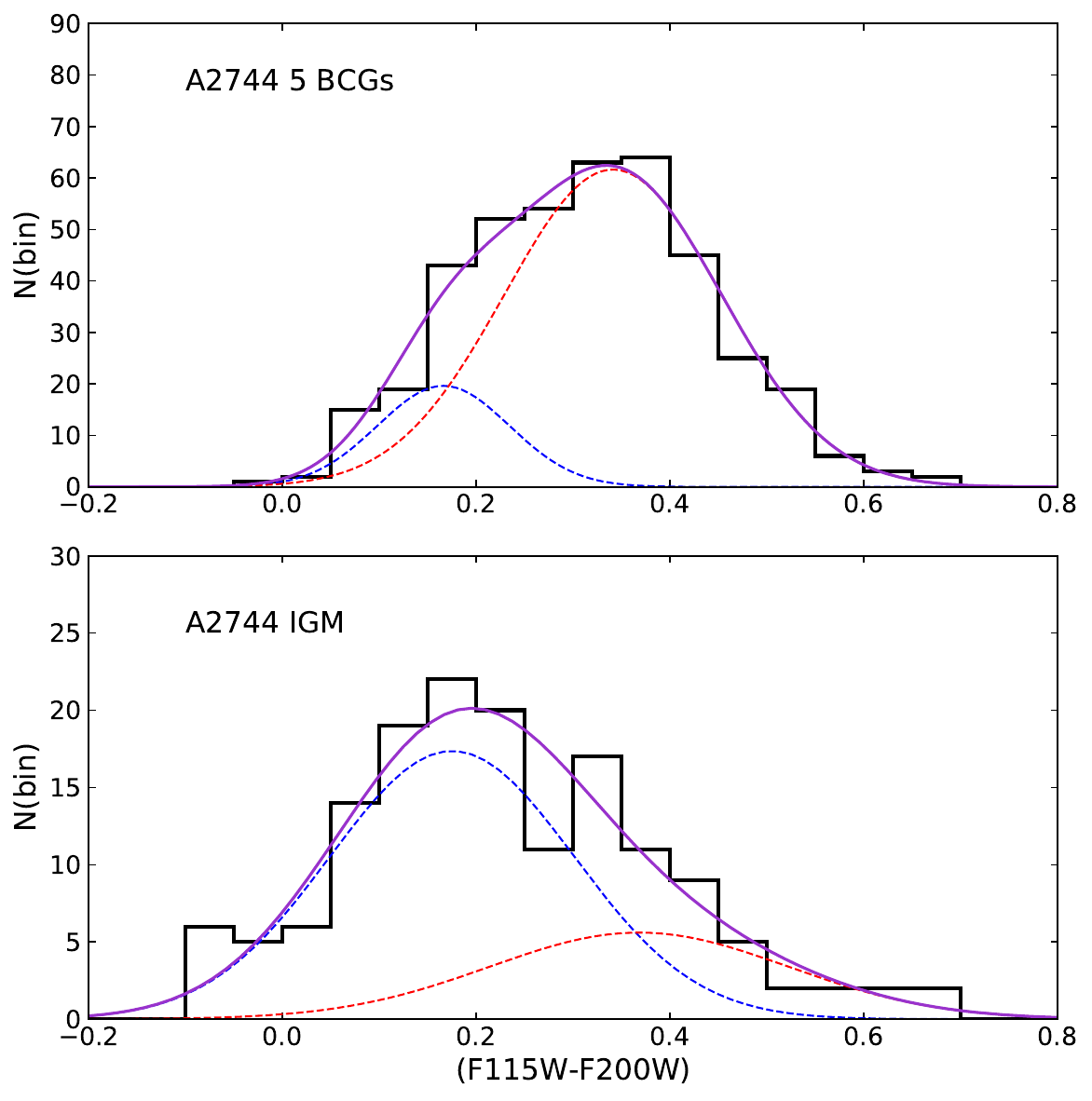}
  \caption{\emph{Upper panel:} CDF for the objects inside the exclusion circles in Fig.~\ref{fig:xypair}, mostly belonging to the five major galaxies in the system.  As in Fig.~\ref{fig:cdf} above, this sample is restricted to the range $27.0 \leq F50W \leq 28.5$. The solid line shows the best-fit double Gaussian solution described in the text, while the dashed blue and red lines show the blue and red subcomponents.  \emph{Lower panel:} CDF for the objects outside the exclusion circles. } \label{fig:cdf_pair}
\end{figure}

In Fig.~\ref{fig:cmd}, just to the left of the major GC sequence, a thin sequence of bluer objects can be seen, in the colour range $-0.3 < (F150W-F200W) < 0.0$ and magnitude range $25 < F150W < 27$.  Close inspection of these objects on the images shows that almost half of them are in very distant background galaxy pairs or groups, and some others are located in the bright bulge light of large galaxies. Some of them may be compact star-forming regions. About 20 of them are isolated, unresolved objects scattered around the field without showing any obvious connection with the A2744 galaxies; their distribution across the field is shown in Figure \ref{fig:xyblue}. If guilt by association is relevant, then these may simply represent other background objects such as remote, high$-z$ galaxy nuclei that were able to pass through the photometric culling steps. Bright, unresolved sources sitting to the blue side of the main GC sequence might also be young massive star clusters that would mark much more recent star formation, such as those seen in the embedded filamentary structure in the  Perseus giant NGC 1275 \citep{holtzman+1992,canning+2014,lim+2020}. However, we find no noticeable numbers of such young blue GCs in the current A2744 data.

The bright end of the CMD distribution extends upward to high luminosity particularly on the red side of the CDF. The spatial distribution of the luminous, red objects $(F150W < 27.5)$ is shown as well in Fig.~\ref{fig:xyblue}. The majority of these objects group closely around the major galaxies, consistent with their being part of the GC/UCD population (see \S\ref{sec:discussion} below for further discussion).

\section{The InterGalactic Medium}

Assuming that GC colour represents metallicity, then an additional test can be made in terms of their spatial distribution. Much previous photometry for GC systems in large galaxies shows that the redder, more metal-rich GCs have more concentrated, steeper radial profiles around their parent galaxies than do the metal-poor clusters \citep{zinn1985,geisler+1996,bassino+2008,faifer+2011,liu+2011,durrell+2014,harris+2017}.  
In Figure \ref{fig:xypair}, the sample has been subdivided into a `blue' group with $(F115W-F200W) < 0.31$ (the midpoint of the CDF as described above), and a `red' group redder than that dividing line. Only objects with $F150W < 29.0$ are included to guarantee high completeness. As seen in the figure, the bluer and redder clusters do not show strong differences in their distributions around the major galaxies. A more noticeable 
difference is found for the objects  scattered across the IGM and in the smaller satellite galaxies, where the bluer objects are much more populous than are the redder ones. 

To demonstrate this latter effect more clearly, the objects near the five major galaxies (BCGs) are marked out by the enclosing circles shown in Fig.~\ref{fig:xypair}.  The resulting CMDs divided by location are shown in Figure \ref{fig:cmd_igm}, which shows the objects outside the circles in the left panel, and the ones inside the circles in the right panel.  For the outer population which consists mainly of GCs in the IGM and a few small satellite galaxies, the red sequence extending to high luminosity ($F150W$ $ < 28$) is almost entirely gone, and by default the blue component is the majority of what remains. 

In Figure \ref{fig:cdf_pair}, the CDFs for both populations are shown, for comparison with Fig.~\ref{fig:cdf}.  A double-Gaussian fit done with GMM as described above yields the results summarized in Table \ref{tab:gmm}.   Within the uncertainties of the fit, the blue and red mean colours ($\mu_1, \mu_2$) and their dispersions ($\sigma_1, \sigma_2$) are similar in all three subsamples listed there. The principal difference is the very much larger blue fraction for the IGM sample.  Similar patterns have been found for the GC populations in the major nearby clusters of galaxies including Virgo \citep{durrell+2014}, Coma \citep{peng+2011}, Fornax \citep{cantiello+2020}, and Perseus \citep{harris+2020}.

The objects in the BCG sample (the ones inside the three exclusion circles) make up 88\% of the total numbers in our photometry.  A very rough upper limit on the IGM population of GCs in A2744 is then that it makes up at most 12\% of the total.  Notably, \citet{montes_trujillo2014}  find that the IntraCluster Light (ICL) in Abell 2744 makes up $>$6\% of the total stellar mass, \citet{morishita+2017} find an ICL fraction of $8-9$\%, while \citet{jimenez-teja+2018} estimate a higher value of $\simeq 16-20$\%. Our result from GC counts favors the two lower values, but the range in these different studies appears to reflect the strong observational challenges in measuring such a wide-field, low-surface-brightness component \citep[e.g.][]{furnell+2021}.  For comparison, \citet{durrell+2014} estimate that in Virgo the IGM GCs make up 18\% of the total in the entire system, while \citet{peng+2011} find that for Coma, the IGM population is in the range 30-45\% of the total.  This contrast adds to the evidence that A2744 is in an earlier stage of its dynamical history than these local galaxy clusters.  At higher redshift, the intergalactic stellar population created by tidal stripping from the member galaxies is still actively developing, and the period since redshift $z \sim 1$ is expected to be especially important for growing the IGM, including both intergalactic GCs and more generally the ICL \citep[e.g.][]{burke+2012,ramos+2015,ramos-almendares+2018,deoliviera+2022,ahvazi+2023}.

In some contrast, for the cluster Abell 1689 at $z=0.183$, \citet{alamo-martinez+2017} find that the intergalactic GCs represent as much at 35\% of the total GC population.  This cluster appears to be in a more dynamically evolved state than A2744, since it has a single dominant central BCG with an extensive cD envelope. Similarly, in the $z=0.39$ cluster SMACS J0723.3-7327, \citet{lee22} find a very extended IGC distribution with a regular, elongated structure centered on the single dominant BCG, which is again suggestive of a fairly advanced evolutionary state.


\section{Luminosity Function}\label{sec:luminosity-function}

For normal old GCs, the luminosity function (GCLF) follows a lognormal shape with a peak $m_0$ and dispersion $\sigma$ that are known to vary slowly with host galaxy mass \citep{jordan+2007,villegas+2010,harris+2014}. The GCLF peak (turnover) point for nearby galaxies is near $M_I \simeq -8.3$ and the dispersion $\sigma \simeq 1.2$ mag \citep{harris+2014}, but the turnover luminosity and dispersion both increase slightly with galaxy mass. In the A2744 galaxies, the apparent magnitude $m_0$ of the GCLF turnover point is expected to be fainter than our photometric detection limit, but the data penetrate very much deeper than did the previous \hst photometry and it should therefore be possible to make a basic test of the GCLF shape.

Figure \ref{fig:lf} shows the observed numbers of objects as measured in $F150W$.  The basic lognormal shape provides a valid description for the brighter magnitude range, but for $F150W \gtrsim 29$, the recovery completeness starts having an important effect on reducing the curve amplitude. For the binned counts shown in Fig.~\ref{fig:lf}  as grey datapoints, we fit the function
\begin{equation}
N(m) = \frac{N_{\rm tot}}{\sqrt{2 \pi} \sigma} e^{-(m-m_0)^2/{2 \sigma^2}} \cdot f(m)
\label{eq:lf}
\end{equation}
where $f(m)$ is the photometric completeness function given above, $N(m)$ is the observed number of objects per unit magnitude in each bin, and $(N_{\rm tot}, m_0, \sigma)$ are the free parameters to be solved for.  An unweighted non-linear least-squares fit gives $m_0 = (31.05 \pm 0.09)$ mag for the turnover point, $\sigma=(1.25 \pm 0.03)$ mag for the dispersion, and a total population $N_{\rm tot} = 78000 \pm 6000$ over the entire system.  The solid magenta curve in Fig.~\ref{fig:lf} shows the intrinsic lognormal LF, while the dashed curve shows the same lognormal function multiplied by the completeness function.

The quoted errors on the three parameters reflect only the internal fitting uncertainties, and are relatively small because the sample size here is large ($10^4$ objects), and as shown in Fig.~\ref{fig:lf}, the data match the numerical model of Eq.~(\ref{eq:lf}) extremely closely.  An unweighted fit avoids giving too much emphasis to either the bright end (where the photometry is higher quality but the bin counts are low) or the faint end (where the counts are far larger but the photometry is more uncertain).  A better idea of the true external uncertainty of the parameters can be gauged if we weight the datapoints by either the absolute count uncertainty $\pm N_{\rm bin}^{1/2}$ or the relative uncertainty $\pm N_{\rm bin}^{-1/2}$.  These give results different by an additional $\pm 0.1$ mag in $m_0$, $\pm 0.05$ mag in $\sigma$, and $\pm 10000$ in $N_{\rm tot}$ compared with the unweighted fit.

Though the lognormal model in Eq.~\ref{eq:lf} fits the observed LF quite well, if the data fall well short of the true GCLF turnover point, then the solutions for $m_0$ and $\sigma$ become correlated \citep{hanes_whittaker1987,harris+2014} and both are increasingly uncertain. At any magnitudes fainter than $F150W \gtrsim 30$, the observed counts are steeply choked off by incompleteness (Fig.~\ref{fig:f}), and all estimated quantities including the completeness function itself become very uncertain.  

The turnover level of $m_0 = 31.05$ predicted from the fit corresponds to an absolute magnitude $M_{F150W} = -9.84$, which in turn converts to a mass of $\simeq 10^6 M_{\odot}$ using the PARSEC models mentioned above.  This level is a factor of $\sim3$ higher than the normal
GCLF turnover at $L \simeq 10^{5.24} L_{\odot}$ established from nearby galaxies \citep{jordan+2007,villegas+2010,harris+2014}, which would lie at $M_{F150W} \simeq -8.9$ or $m_0 \simeq 32.0$ at the distance of A2744.  To check further on the impact of changing the LF parameters, the fit can be redone by forcing either the dispersion $\sigma$ or the turnover $m_0$ to take fixed values closer to the ones initially expected.  If we adopt $\sigma \equiv 1.4$ mag as an appropriate mean for large galaxies \citep{harris+2014}, the resulting two-parameter solution gives $N_{\rm tot} = 116000, m_0 = 31.54$.  Similarly, if we adopt $m_0 \equiv 32.0$ as a normally expected turnover luminosity, the resulting solution gives $N_{\rm tot} = 167000, \sigma = 1.52$ mag. All three sample solutions are shown in Fig.\ref{fig:lf},  (dashed magenta line for the three-parameter solution, dashed blue line for fixed $m_0=32.0$, dotted green line for $\sigma=1.4$).  There is little to choose among them aross most of the range, though the restricted two-parameter solutions tend to overpredict the numbers of the brightest clusters. Arguments can be made for any of these solutions, and we conclude for the present only that the total GC population is likely to be at least $\sim 10^5$ clusters.  We also cannot rule out a GCLF turnover or intrinsic dispersion close to the standard values for giant galaxies.  Most importantly, we view this comparison primarily as a test that the standard lognormal LF shape provides a reasonable match to the data.

It should also be emphasized that the best-fit parameters $(m_0, \sigma)$ are an average over the entire A2744 system.  The lookback time for A2744 is large enough that some evolution of the GCLF shape should be expected. Simulations of evolving GC systems within Milky-Way-sized galaxies with EMP-\textit{Pathfinder} \citep{reina-campos+2022b} show that by an age of $\sim 6-7$ Gyr, the basic lognormal shape of the GCLF is well established, but the peak point and spread (dispersion) continue to evolve slowly as the clusters age and disrupt, in the sense that the GCLF turnover will gradually become brighter and the spread narrower \citep{kruijssen2009,choksi_gnedin2019,reina-campos+2022b}.

An additional factor to be considered in a more detailed analysis would involve the effects of differential ages. The GCs in the Milky Way show an age-metallicity relation \citep{vandenberg+2013,forbes_bridges2010,dotter+2011} in the sense that the metal-poor ones are  12-13 Gyr old while the most metal-rich are 9-10 Gyr old (with scatter). Thus if the A2744 GCs have a similar age-metallicity relation, then at a lookback time of 3.5 Gyr the metal-poor ones are seen as $\simeq9$ Gyr old while the metal-rich ones are only $\simeq6$ Gyr old; i.e.~the relative age difference was larger at that time. The red sequence should then be relatively brighter than the blue sequence for the same GC mass range.  A hint of that effect may be visible in the CMD, where the redder side of the distribution reaches to a much higher level (cf.~the discussion below). Further investigation into these effects will be done in later work. Deeper photometry would reveal a much clearer picture, but that is not likely to be achievable in the foreseeable future.

The estimated total GC population from the three-parameter solution of $N_{\rm tot} \simeq 78000$, as stated above, includes all the cluster galaxies and the IGM over the central region.  It should be viewed as a lower limit, since it does not include regions outside the central part of the cluster studied here, and does not account for other locations of area incompleteness such as near the centers of the large galaxies. This total is nevertheless smaller than the estimate of $4 \times 10^5$ GCs from \citet{lee16} from the \hst imaging, though later discussions \citep{harris+2017,alamo-martinez+2017} revised that downward by at least a factor of two. These \hst-based estimates were based on only the brightest magnitude or so of the GCLF and thus required a very large extrapolation compared with our present data. {As more relevant comparisons, our estimate for A2744 from the basic three-parameter GCLF fit for $N_{\rm tot}$ is similar to the total of 67000 GCs in the entire Virgo cluster \citep{durrell+2014}, and $\simeq 100000$ GCs in the Coma cluster \citep{peng+2011}, again including both the IGM and the member galaxies.  Notably, however, either of our two-parameter GCLF solutions stated above would raise $N_{\rm tot}$(A2744) above $10^5$, appropriate for the large total mass of the system \citep{harris+2017b,dornan_harris2023}.

\begin{figure}
	\includegraphics[width= 0.90\columnwidth]{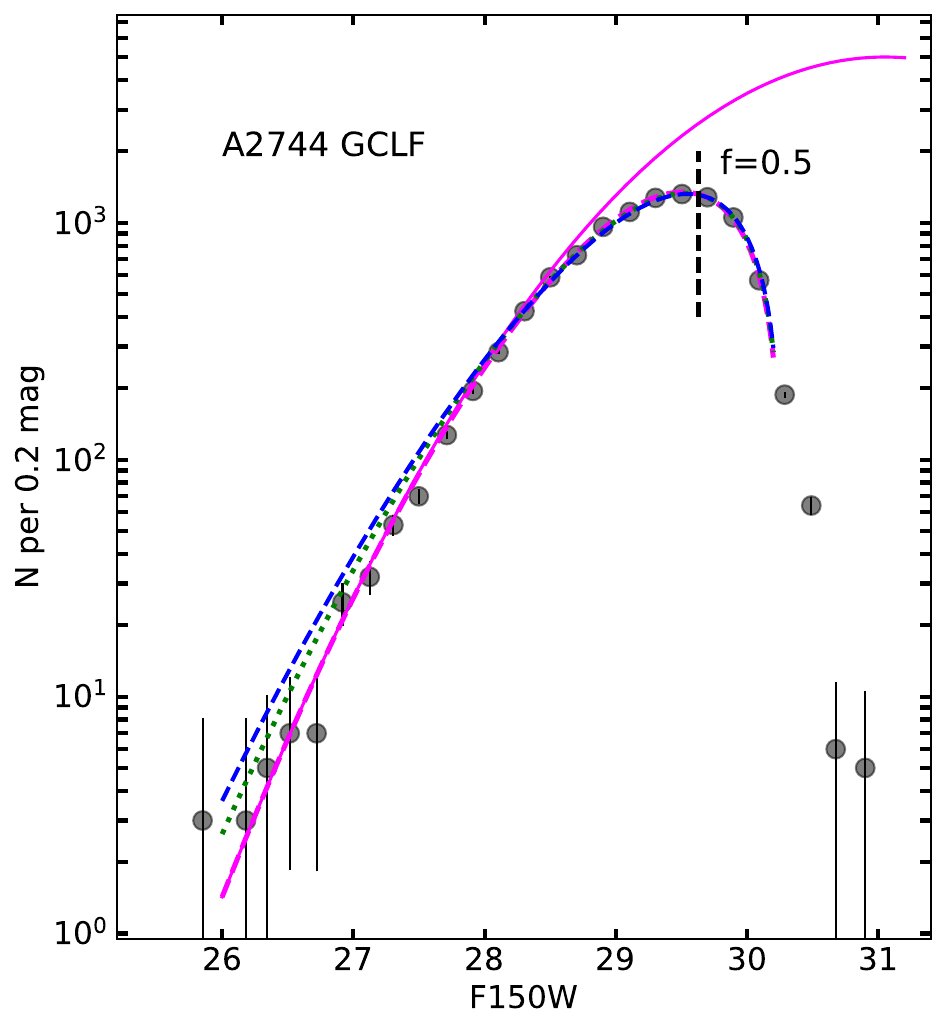}
    \caption{Luminosity function for the GCs in the A2744 field. The gray datapoints and errorbars show the observed number of objects per $0.2-$mag bin in $F150W$, with the $50~$per cent recovery completeness level as indicated by the vertical dashed line. The rising solid line (magenta) shows the best-fit three-parameter solution for the GCLF with $m_0=31.05, \sigma=1.25$, while the three curves going through the datapoints as described in the text show three possible solutions for the GCLF now multiplied by the photometric completeness function. Fainter than $F150W \gtrsim 30$, all numbers are very uncertain.} \label{fig:lf}
\end{figure}

\begin{figure}
	\includegraphics[width=\columnwidth]{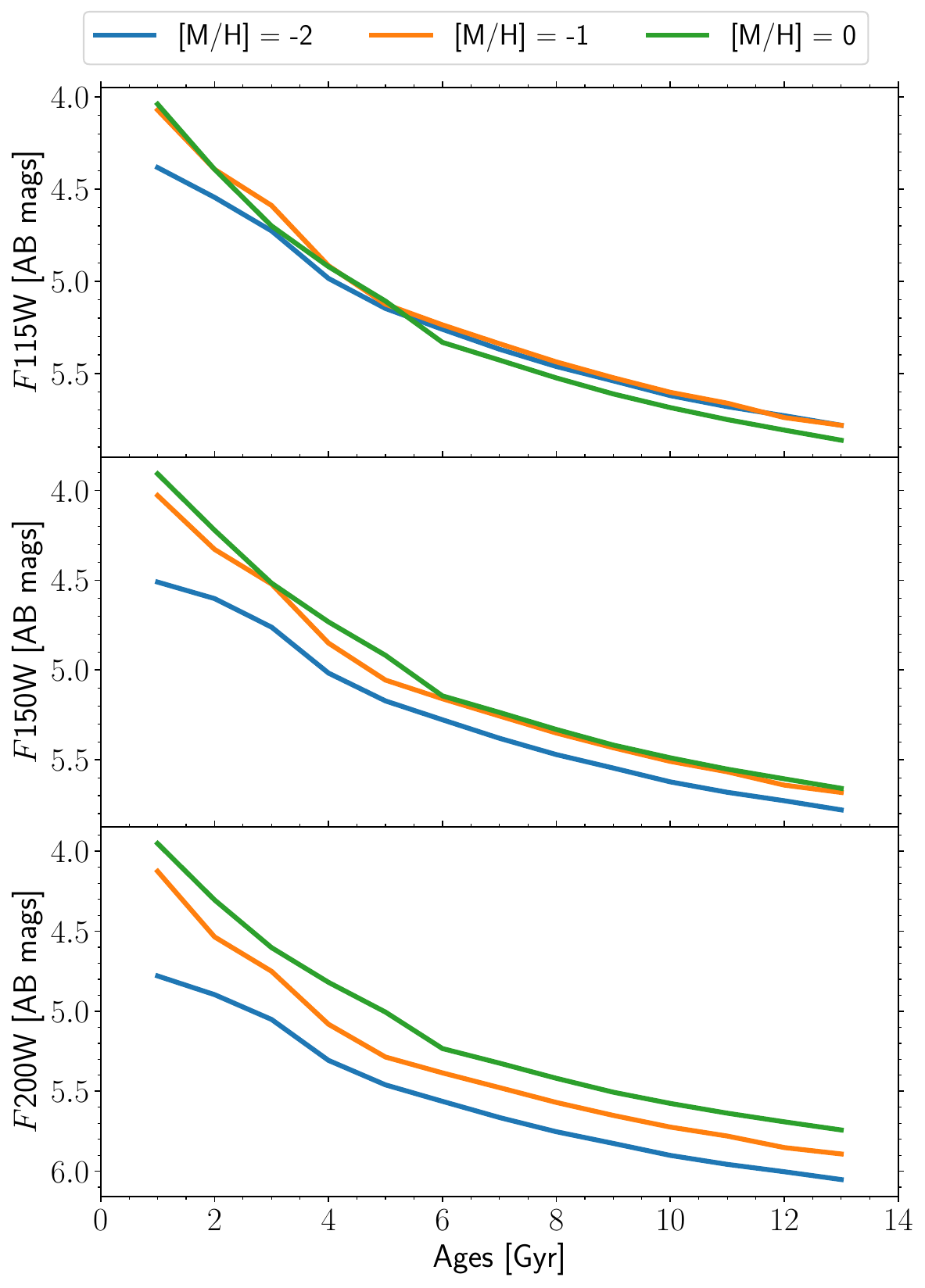}
    \caption{Predicted GC luminosity versus age in the three NIRCAM filters. Blue lines are for [M/H] $= -2.0$, orange for $-1.0$, and green for 0.0. These are calculated from the PARSEC CMD3.7 stellar models as described in the text.  These curves track the passive decline in luminosity due only to stellar evolution, and do not include the effects of mass loss (see text). The absolute magnitudes are normalized to $1 M_{\odot}$.} \label{fig:agetrend}
\end{figure}

\section{Adjustments to Zero Redshift}\label{sec:adjustments-z0}

Comparison of the A2744 GC population with local ($z=0$) GCs requires estimating the size of three different effects arising from the lookback time and redshift.

\subsection{Passive Luminosity Evolution} 

Any star cluster after its initial formation period will decrease in luminosity over time due to simple stellar evolution. In Figure \ref{fig:agetrend}, predicted luminosities in the three NIRCAM filters as derived from the PARSEC models are shown, at three different metallicities. Although the absolute magnitude depends slightly on metallicity, the relative change in luminosity over the past 3.5 Gyr is nearly independent of metallicity or filter. The net change from $z=0.3$ to $z=0$ is an age-fading of $\simeq 0.21$ mag, or about $20~$per cent decrease in luminosity.

This estimate, however, assumes a similar mean age for all the GCs. If as noted above the red GCs are as much as 3 Gyr younger on average than the blue GCs, then they would decrease by about $30~$per cent in luminosity down to the present time. A mean age difference of $\sim 3$ Gyr may be an upper limit, however, since the age/metallicity relation (AMR) for GCs is expected to be steeper for more massive galaxies \citep{horta+2021}, and most of the A2744 GCs are clearly associated with the giant member galaxies.

\subsection{Mass Loss} \label{subsec:mass-loss}

All GCs experience progressive mass loss due to stellar evaporation and tidal stripping. Clusters located in the IGM far from any major galaxies will be subject to only slow stellar evaporation. However, most of the GCs in our observed sample are satellites of the major galaxies (Fig.~\ref{fig:field}b) and these will experience larger tidal losses. Tidal disruption times for GCs will be several Gyr \citep[e.g.][]{baumgardt_makino2003} but the mass loss rate depends on their own mass and their local host-galaxy potential.  A useful first-order prescription for the fractional mass loss, averaging over galactocentric radius, is \citep{choksi_gnedin2019}
    \begin{equation}
    \frac{\Delta M}{M} = -\frac{\Delta t}{t_{\rm tid}} \simeq -\frac{\Delta t}{\textrm{5 Gyr}} \cdot \left(\frac{M}{2 \times 10^5 M_{\odot}} \right)^{-2/3}
    \end{equation}
where $t_{\rm tid}$ is the tidal disruption time and the GCLF turnover mass is $2-3 \times 10^5 M_{\odot}$ \citep{villegas+2010,harris+2014}.  \citet{baumgardt_makino2003} or \citet{lamers+2005} give similar relations. As discussed above, the A2744 GCs  observed in this study are on the bright half of the the GCLF, with masses extending to $\simeq 10^6 M_{\odot}$ and above at the bright end, so with $\Delta t = 3.5$ Gyr, the ratio ($-\Delta M/M$) will range from $\simeq 0.15$ at the high-mass end up to as much as 0.7 at the GCLF peak; i.e. a dimming of 0.2 to 1.3 mag in absolute magnitude.  Stellar evaporation is negligible by comparison, with dissolution times one or two orders of magnitude larger \citep[e.g.][]{baumgardt_makino2003,lamers+2005}.

\subsection{K-Corrections}\label{subsec:k-correction}

The cosmological K-correction is well known in galaxy studies but relatively unfamiliar for GC work. The redshift causes any rest-frame wavelength $\lambda_0$ emitted at the source to be measured at $\lambda = (1+z)\lambda_0$, thus both shifting and stretching any given filter bandpass and reducing the energy of each photon by the same factor. By definition, the absolute magnitude $M_{\lambda}$ at wavelength $\lambda$ is
\begin{equation}
    M_{\lambda} = m_{\lambda} - 5 \log_{10}\left( \frac{d_L}{10\,{\rm pc}} \right) - K_{\lambda}  ~.
    \label{eq:kcorr} 
\end{equation}
The redshift factor $K_{\lambda}$ may be positive or negative depending on the shape of the spectrum in the region of a given filter bandpass. For A2744, there is an interesting coincidence at play: the ratios of effective wavelengths between the adjacent NIRCAM/SWC filters almost exactly match its redshift factor of $(1+z) = 1.3$. That is, light observed in $F200W$ ($\lambda_{\rm eff} = 19680$\AA) was emitted very nearly in $F150W$ (14873 \AA); light observed in $F150W$ was emitted in $F115W$ (11434 \AA); and light observed in $F115W$ was emitted in $F090W$ (8985 \AA).  Thus for the colour indices, our most metallicity-sensitive index $(F115W-F200W)$ is actually close to the rest-frame $(F090W-F150W)$, a colour that is similarly metallicity-sensitive.

GCs are good approximations to single stellar populations (SSPs) and their spectral energy distributions (SEDs) are not the same as those of galaxies. For galaxies, the $K$-values at a given redshift and wavelength depend on galaxy type (essentially, star formation history) \citep[e.g.][]{blanton_roweis2007,o'mill+2011,beare+2014}, whereas for SSPs the important parameters are instead age and metallicity. A more general treatment of the problem is therefore of interest, and this topic is developed further in a separate discussion (Reina-Campos \& Harris 2023, in preparation). For the purposes of this first look at the A2744 data, we apply mean $K$ corrections for only a single age and metallicity.

To estimate the $K$-values, model template SEDs were first generated with E-MILES\footnote{http://miles.iac.es} \citep{rock+2016}.  These have been used to calculate $K_{\lambda}$ for the three bandpasses here following the method outlined in \citet{hogg+2002,blanton_roweis2007,condon_matthews2018}. For the SEDs, BaSTI isochrones were adopted with an age of 9 Gyr, metallicity [Fe/H] = $-1$, and a Chabrier IMF. With these assumptions we obtain $K_{115} = -0.17$ mag, $K_{150} =-0.17$, and $K_{200} =-0.42$. The values for $F115W$ and $F150W$ are the same because the redshifted and rest-frame SEDs are virtually parallel across those two bandpasses. Use of PARSEC isochrones by comparison gave negligible differences.  The true uncertainties in these estimates are hard to gauge since they rely on model SEDs and also depend on the adopted age and metallicity, but are likely to be at least $\pm0.05$ mag \citep[cf. the discussion of][]{blanton_roweis2007}.

\begin{figure}
	\includegraphics[width=\columnwidth]{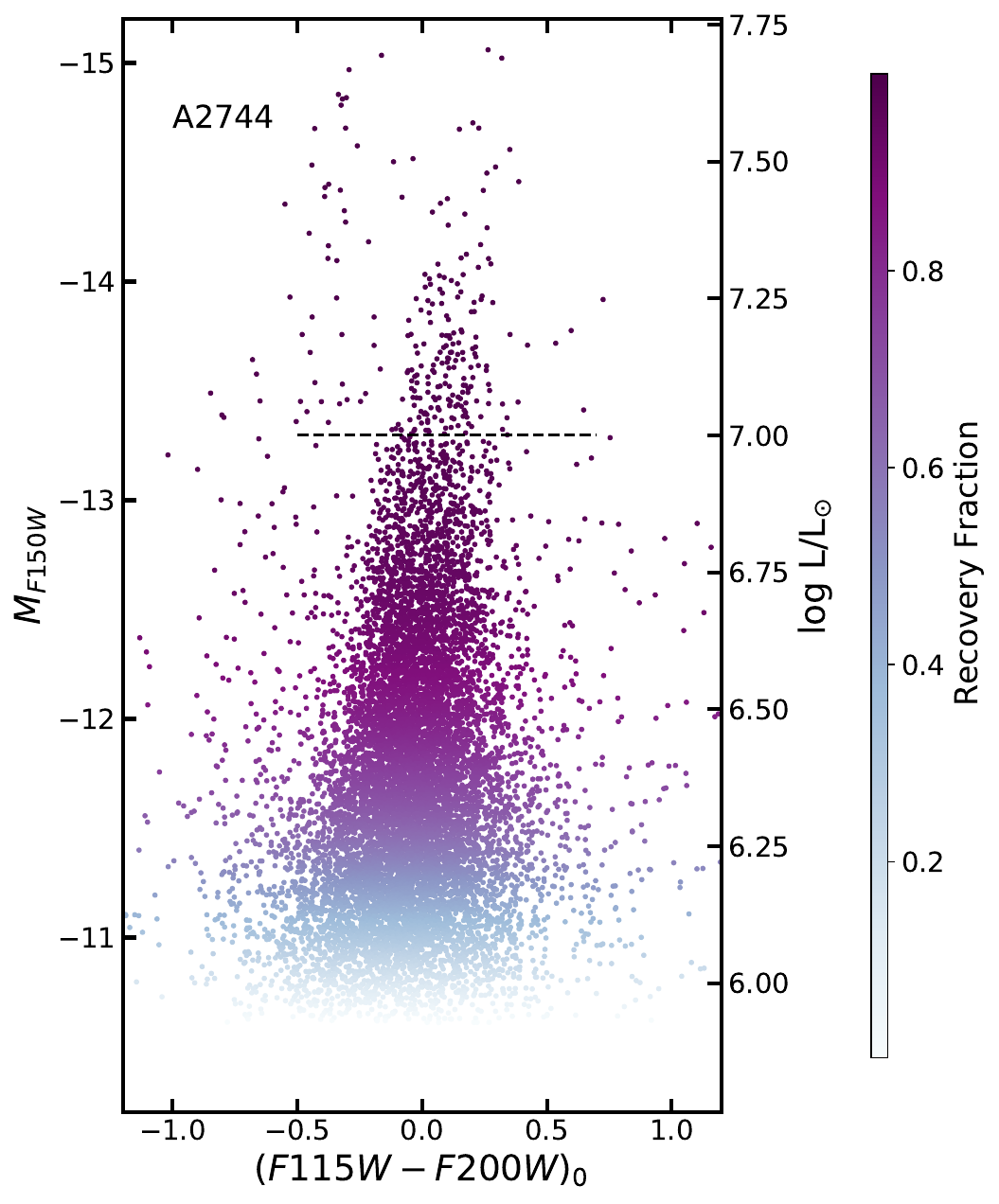}
    \caption{Absolute magnitude versus intrinsic colour for the GC population in A2744. Points are colour-coded by recovery probability as determined from the artificial-star tests. Luminosities and colours include the $K$-corrections. The dashed line indicates the approximate UCD luminosity threshold corresponding to a luminosity of $\simeq10^7 L_{\odot}$ (see text).} \label{fig:cmd_prob}
\end{figure}

\section{A UCD Population}\label{sec:discussion}

To the measured magnitude in each filter, we derive the absolute magnitude from Eq.~\ref{eq:kcorr} with the appropriate $K$-correction and $d_L = 1630$ Mpc.  
In Figure \ref{fig:cmd_prob}, the CMD for the system is shown again, but in the form of luminosity versus intrinsic colour.


The CMD of Fig.~\ref{fig:cmd_prob} indicates, as mentioned previously, that the luminosity distribution of our measured objects reaches to very high levels and into the UCD regime.  UCDs have generated an extensive literature since their discovery in the Fornax cluster \citep{hilker+1999,drinkwater+2000}: comprehensive lists of references are given, for example, by \citet{misgeld+2011,penny+2012,zhang+2015,lee16,janssens+2017} and our current discussion can mention only a few. See also \citet{hilker2009,misgeld_hilker2011,forbes+2014,norris+2014,janz+2016} for overviews of how UCDs fit within the broader context of hot stellar systems including GCs, cE's, E galaxies, and nuclear star clusters and the overlaps among all these categories in terms of scale size versus luminosity or mass.  At least three different plausible formation mechanisms have been debated including  tidal stripping of compact dwarfs or extensions of the GC sequence \citep[see, e.g.,][as well as the references cited above]{mieske+2008,evstigneeva+2008,mieske+2012,lee16} and the UCD population seen in total may be a heterogeneous combination of all these channels.

UCDs around nearby galaxies have traditionally been identified through combinations of properties including high luminosities ($M_{\rm V} \lesssim -11$), large scale radii ($r_{\rm eff} > 10$ pc), and internal dynamics ($M/L \gtrsim 2$) that set them apart from normal GCs.  There is no universal definition of a luminosity or mass threshold, for example, below which a compact stellar system is a GC and above which it is a UCD; that is, the GC and UCD regimes overlap (cf.~the references cited above).  \citet{janssens+2017,janssens+2019} extract a UCD sample for A2744 from the \hst Hubble Frontier Field imaging data, using stellar models to convert $I-$band luminosity to mass along with a threshold $M = 2 \times 10^6 M_{\odot}$ to set the lower limit of the UCD regime \citep[see also][for adoption of the same threshold]{mieske+2008}. Other studies have used various luminosity-based thresholds typically in the range $\sim 10^{6-7} L_{\odot}$ \citep[e.g.][]{mieske+2006,chiboucas+2011,harris+2014,liu+2015,evstigneeva+2008,wehner_harris2007,penny+2012,mieske+2012}.

For A2744, all GCs and UCDs are unresolved and we have only the luminosities and colours to work with. Rather than mass, we prefer to adopt a luminosity-based UCD threshold here to minimize dependence on the stellar models. The LF discussed in the previous Section fits a standard lognormal curve rather seamlessly up to the highest luminosities, so identifying a point at which GCs `end' and UCDs begin is to some degree arbitrary \citep[cf.][]{mieske+2012}. The CMD does, however, show a change in distribution near $M_{F150W} \simeq -13.3$ (Fig.~\ref{fig:cmd_prob}, $F150W \simeq 27.6$ in Fig.~\ref{fig:cmd}), above which the blue-GC sequence essentially disappears and the red sequence continues upward.  This feature is visible in the present \jwst data because of its much higher precision and depth compared with the shallower \hst photometry \citep{lee16,janssens+2017}, though a hint of it can be seen in Figure 6a of \citet{lee16}.  Notably, this same pattern shows up in the GC CMDs for many other giant galaxies where the red sequence continues to distinctly higher levels \citep{mieske+2006,wehner_harris2007,harris2023}.  With $M_{150,AB}(\odot) = 4.21$, this level corresponds very nearly to $10^7 L_{\odot}$.  Accounting for $\simeq 0.4$ mag of luminosity evolution and mass loss over 3.5 Gyr (see above) would bring the transition point to $7 \times 10^6 L_{\odot}$ at redshift $z=0$, consistent within factors of two or three with many of the UCD studies mentioned above.  As seen in Fig.~\ref{fig:xyblue}, most of these very luminous objects concentrate quite closely around the five major galaxies in the system \citep{janssens+2017,janssens+2019}, a result that has also been found for UCDs in nearby clusters such as Virgo and Coma \citep{chiboucas+2011,liu+2015}. 

In Fig.~\ref{fig:cmd_prob}, there are 200 objects higher than the luminosity threshold stated above, which can plausibly be called UCD candidates.  But their scale sizes, M/L ratios, and ages are unknown, so these may also be a mixture of UCDs and normal GCs.  For comparison, \citet{janssens+2017} found N(UCD) $= 385 \pm 32$ in the \emph{mass} range $10^7-10^8 M_{\odot}$, not specifically corrected for detection incompleteness.  Mass-to-light ratios are uncertain and can differ between UCDs, but for a mean M/L = 4 \citep[e.g.][]{mieske+2008} this range would correspond to $L \gtrsim 2.5 \times 10^6 L_{\odot}$, lower than the transition adopted here.  \citet{lee16} concluded N(UCD) $= 147 \pm 26$ for $F814W(\rm Vegamag) < 28.6$ ($F150W$ $\lesssim 28.7$) by explicitly subtracting off a GCLF with an assumed peak and dispersion from the LF of the total sample, though measured from a much smaller region than ours.  Given the different methods, area coverage, and adopted transition points, these studies appear to be at least roughly consistent.

\section{Summary and Conclusions}\label{sec:summary}

In this study, we have used the deep UNCOVER mosaic images of Abell 2744 taken with \jwst NIRCAM to construct a catalog of 10,000 pointlike (unresolved) objects in the central field of the cluster. The vast majority of the detected objects consists of globular clusters populating the halos of the A2744 galaxies, but many of them also scatter throughout the InterGalactic (IntraCluster) Medium. PSF fitting with the tools in \texttt{daophot} has been used to complete the photometry in the $F115W$, $F150W$, and $F200W$ filters, in all of which the photometric limits are closely similar. Extensive artificial-star tests are used to determine the recovery completeness and internal measurement uncertainties.  

A brief summary of our findings is as follows:
\begin{enumerate}
    \item[(1)] After careful rejection of nonstellar or crowded sources, the remaining sample of starlike (unresolved) sources defines a sharp, well populated sequence in the CMD. This sample is dominated by GCs, with very little residual contamination.
    \item[(2)] The spatial distribution of the detected sources shows that the majority are located within the halos of the five biggest A2744 galaxies. The map of their distribution also marks out many satellites, but there is also a significant population distributed more widely across the IGM.
    \item[(3)] The colour distribution particularly in $(F115W-F200W)$ is much wider (0.5 mag) than the $\lesssim 0.2-$mag spread that would be expected from measurement scatter alone, and is what would be expected if the GCs span the full usual range of metallicities from [Fe/H] $\simeq -2$ to 0. There is no indication of star clusters that were formed much more recently, which would show up as a population of bright clusters much bluer than the normal GC sequence.
    \item[(4)] The GC luminosity function covers a $2.5-$mag range before it is severely damped by photometric incompleteness. Over this range, however, it accurately matches the lognormal shape that is expected for normal old GCs.  Extrapolating to fainter magnitudes, we estimate the total GC population in the region surveyed here to be $\simeq 10^5$ or more objects.  
    \item[(5)] The metal-poor GCs (defined as the bluer half of the colour range) are less centrally concentrated to the galaxies than are the metal-richer GCs, and dominate the IGM population.
\end{enumerate}

The present data strongly suggest that in A2744 we are looking at classic, normal populations of GCs around its member galaxies, seen at an earlier stage but well on their way to evolving into the types of distributions that we see in the Local Universe. Already by this epoch, the lognormal shape of the LF, at least at the high-luminosity end, has been well established, as have their fairly regular distributions around the halos of the member galaxies. The colour distribution function, after adjustment for K-corrections, reflects that of the full expected metallicity range for old GCs.

Much more detailed followup work is possible and will be addressed in later studies. Normalizations to the data need to be improved, including $K$-corrections as a function of colour (metallicity) and age, luminosity evolution, and dynamical mass loss as a function of both GC mass and spatial location. Quantitative derivation of the GC radial profiles around the major A2744 galaxies needs to be done as well. Such work will lead to more detailed comparison of the overall GC distribution with other indicators of the DM distribution including the gravitational lensing map and the X-ray gas.

Above all, the A2744 data provide an excellent example of the new ground opened up by \jwst for the study of rich clusters of galaxies and their embedded GC systems. For the first time, it is possible to investigate observationally the evolution of entire GC systems over the past several Gyr leading up to the present day.

\section*{Acknowledgements}

We express our sincere thanks to Laura Parker, John Weaver, Laura Greggio, and Jeremy Webb for helpful discussion, and to Jose Diego for pointing us to this dataset.  The anonymous referee made several helpful suggestions for revisions and additions to the discussion.  We would also like to acknowledge the work of the UNCOVER team to produce the beautiful mosaic images and to make them publicly available at such an early stage. 
MRC gratefully acknowledges the Canadian Institute for Theoretical Astrophysics (CITA) National Fellowship for partial support. This work was supported by the Natural Sciences and Engineering Research Council of Canada (NSERC). 

\emph{Facility:} JWST (NIRCAM)

\emph{Software:} IRAF \citep{tody1986,tody1993}, Daophot \citep{stetson1987}, Matplotlib \citep{hunter2007}, E-MILES \citep{rock+2016}, PARSECv1.2 \citep{bressan+2012}, BaSTI \citep{pietrinferni+2021}.

\section*{Data Availability}

The mosaics for Abell 2744 from the UNCOVER program are publicly available at: \href{https://jwst-uncover.github.io/#}{https://jwst-uncover.github.io/\#}.



\bibliographystyle{mnras}
\bibliography{mybib} 

\bsp	
\label{lastpage}
\end{document}